\documentclass{def2}            

\usepackage{natbib1}
\bibpunct[]{(}{)}{,}{a}{}{,}

\begin{document}
\input epsf.def   
\input psfig.sty

\jname{\quad\quad \sl G\"udel M. 2002, Annual Review of Astronomy and Astrophysics {\bf 40}:217-261}
\jyear{}
\jvol{}
\ARinfo{}

\def\ga{\;\rlap{\lower 2.5pt\hbox{$\sim$}}\raise 1.5pt\hbox{$>$}\;}       
\def\la{\;\rlap{\lower 2.5pt\hbox{$\sim$}}\raise 1.5pt\hbox{$<$}\;} 

\title{Stellar Radio Astronomy\\
{\Large Probing Stellar Atmospheres from Protostars to Giants}}

\markboth{G\"udel}{Stellar Radio Astronomy}

\author{Manuel G\"udel
\affiliation{Paul Scherrer Institut, W\"urenlingen \& Villigen, CH-5232
             Villigen PSI, Switzerland\\
	     e-mail: guedel@astro.phys.ethz.ch}}

\begin{keywords}
radio stars, coronae, stellar winds, high-energy particles, non-thermal radiation, magnetic fields
\end{keywords}

\begin{abstract}
Radio astronomy has provided evidence for the presence of 
ionized atmospheres around almost all classes of non-degenerate stars. Magnetically
confined coronae dominate in the cool half of the Hertzsprung-Russell 
diagram. Their radio emission is predominantly of non-thermal origin and 
has been identified as gyrosynchrotron radiation from mildly relativistic 
electrons, apart from some coherent emission mechanisms.
Ionized winds are found in hot stars and in red giants. They
are detected through their thermal, optically thick radiation, but
synchrotron emission has been found in many systems as well. 
The latter is emitted presumably by shock-accelerated electrons in weak 
magnetic fields in the outer wind regions. Radio emission is also frequently 
detected in pre-main sequence stars and protostars, and has recently been 
discovered in brown dwarfs. This review summarizes the radio 
view of the atmospheres of non-degenerate stars, focusing on energy release physics
in cool coronal stars, wind phenomenology in hot stars and cool giants, and 
emission observed from young and forming stars.  
\end{abstract}

\maketitle

{\ } \vskip 0.5truecm

\hbox{
\hskip 3.8truecm \begin{minipage}{12cm}
{\it Eines habe ich in einem langen Leben gelernt, n\"amlich, dass unsere ganze Wissenschaft,
an den Dingen gemessen, von kindlicher Primitivit\"at ist - und doch ist es das K\"ostlichste,
was wir haben.}\\ 
One thing I have learned in a long life: that all our science, measured against
reality, is primitive and childlike - and yet it is the most precious thing we have.\\
{\ }
A. Einstein 1951, in a letter to H. M\"uhsam, Einstein Archive 36-610
\end{minipage}
}

\section{INTRODUCTION}

Stellar radio astronomy has matured over the 
past two decades, driven in particular by discoveries made with the 
largest and most sensitive radio interferometers.  Radio emission
is of great diagnostic value as it contains telltale
signatures not available from any other wavelength regime. Some of the 
detected radio emission represents the highest-energy particle populations 
(MeV electrons) yet accessible on stars, the shortest (sub-second) detectable 
time scales of variability and energy release, and probably refers most 
closely to the primary energy release responsible for coronal heating. 
This review is to a large extent devoted to demonstrating the ubiquity 
of high-energy processes in stars as revealed by radio diagnostics. 

Stellar radio sources include thermal and non-thermal magnetic coronae, 
transition regions and chromospheres, stars shedding winds, colliding-wind binaries, 
pre-main sequence stars with disks and radio jets, and embedded young objects 
visible almost exclusively by their radio and millimeter-wave emission.
Most recent additions to the zoo of objects are brown dwarfs, and
with the increasingly blurred transition from  stars to brown dwarfs to
giant planets like Jupiter and Saturn, even the magnetospheres of the latter 
may have to be considered a manifestation of magnetic activity in the widest 
sense. To keep the discussion somewhat focused,
this paper concentrates on physical processes in magnetic coronae, but 
includes,  in a more cursory way, atmospheres of young and forming stars and
winds of hot stars. We do not address in detail the large phenomenology of extended
and outflow-related sources such as radio jets, HII regions, masers,
Herbig-Haro objects,  and the diverse millimeter/submillimeter phenomenology, e.g., 
molecular outflows and dust disks. Compact stellar objects (white dwarfs, neutron stars, 
black holes) are not considered here. 

Inevitably - and fortunately - much of the knowledge gained in stellar astronomy
is anchored in solar experience. The privilege of having a fine specimen - and
even an exemplary prototype - next door is unique among various fields
of extrasolar astrophysics, being shared since recently only by the related 
field of extrasolar planetary astronomy. Detailed solar studies, even in-situ
measurements of the solar wind, are setting high standards for studies
of stellar atmospheres, with a high potential reward. Solar radio astronomy
has been reviewed extensively in the literature.
For detailed presentations, we refer to \citet{dulk85} and \citet{bastian98}.

The maturity of stellar radio astronomy is demonstrated by a number of  review
articles on various subjects; a non-exhaustive list for further 
reference includes
\citet{andre96, bastian90, bastian96, bookbinder88, bookbinder91, dulk85, dulk87, 
guedel94a,  hjellming88, kuijpers85a, kuijpers89a, lang90, lang94, lestrade97,
linsky96, melrose87, mullan85, mullan89, mutel96, phillips91r, seaquistrev93, vdoord96a}, and 
\citet{white96, white00}.
A natural starting point for this review, even if not strictly adhered
to, is Dulk's comprehensive 1985 Annual Review  article that summarizes the pre- and 
early-Very Large Array view of stellar (and solar) radio emission.
Meanwhile, two dedicated conferences, the first one in Boulder in 1984 (proceedings
edited by \citealt{hjellming85}) and the second held in Barcelona in 
1995 (proceedings edited by
\citealt{taylor96}), provided a rich forum to discuss new developments; 
together, they beautifully illustrate the progress made over the past decades.

\section{RADIO SURVEYS AND THE RADIO HERTZSPRUNG-RUSSELL DIAGRAM}

\begin{figure}[t!]
\centerline{\psfig{file=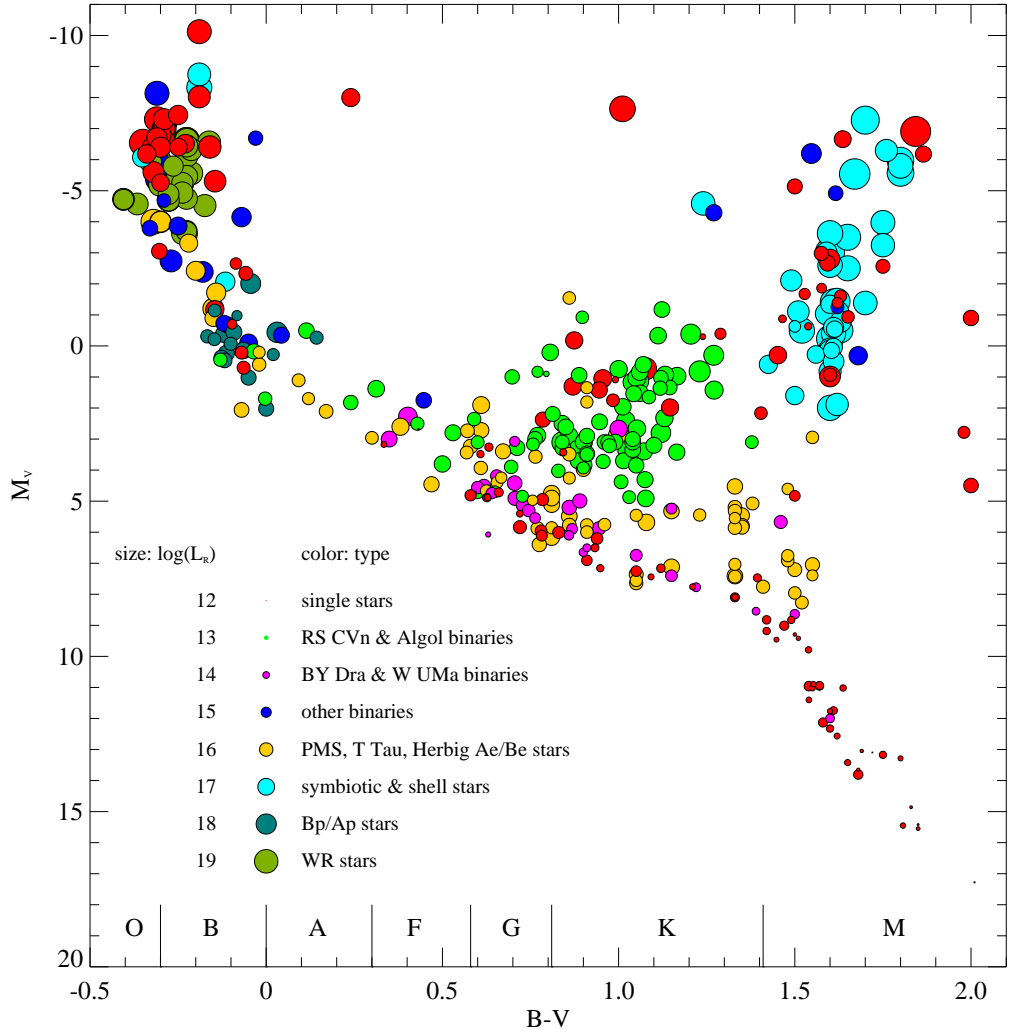,width=14.5truecm}}   
\caption{HR diagram showing 440 radio-detected stars.\label{hrd}}
\end{figure}

Figure~\ref{hrd} presents a radio Hertzsprung-Russell Diagram (HRD) based on stellar radio
detections between 1$-$10~GHz, reported in the catalog of \citet{wendker95}; for other 
examples, see \citet{white96, white00}. The luminosity is the logarithmic average
of all reported detections. The accuracy of the location of some stars on the HRD is 
compromised by the limited quality of distance and color measurements, or in multiple 
systems,  by the uncertain attribution of the radio emission to one of the components.
Nevertheless, almost all  of the usual features of an HRD  are recovered, 
testifying to the importance and ubiquity of radio emission. 

The many nearby M dwarfs in Figure~\ref{hrd} were among the first radio stars surveyed. 
The typically larger distances to earlier-type cool dwarfs made their discovery more of 
an adventure, but the samples now include late-type binaries,  K 
stars, solar analog G stars, and a few F stars.
Most radio-detected dwarf stars are strong X-ray sources and young, rapid rotators. 

The cool half of the subgiant and giant area is dominated by the large and radio-strong
sample of RS CVn and Algol-type close binaries. Other sources in this area include
the vigorous coronal radio sources of the FK Com class, chromospheric 
radio sources, symbiotic stars and  thermal-wind emitters.  Additionally,
a very prominent population of thermal or non-thermal sources just above the main 
sequence is made up of various classes of pre-main sequence stars, such as classical 
and weak-lined T Tauri stars.  

Common to most cool star radio emitters is their non-thermal nature  
which, especially in its ``quiescent'' form, constitutes one of the most
significant - and unexpected - discoveries in stellar radio astronomy, suggesting
the presence of magnetic fields and high-energy electrons. 
Moving toward A-type stars on the HRD, one expects, and
finds, a  dearth of radio detections owing to the absence of magnetic dynamo action. 
However, this is also the region of the chemically peculiar Ap/Bp stars that possess strong
magnetic fields and many of which are now known to be non-thermal  radio sources as well.
Some pre-main sequence Herbig Ae/Be stars in this area are also
prominent radio emitters.

Finally, the hot-star region is heavily populated by luminous
wind-shedding O and B stars and Wolf-Rayet (WR) stars, both classes showing
evidence for either thermal or non-thermal radio emission.

\section{THEORY OF RADIO EMISSION FROM STELLAR ATMOSPHERES}

\subsection{Elementary Formulae for Radiation and Particles}

We summarize below some handy formulae that
have become standard for most radio-stellar interpretational work
(the interested reader
is advised to consult the original literature as well).
In thermodynamic equilibrium, the emissivity of a plasma of temperature 
$T_{\rm eff}$, $\eta_{\nu}$ (in erg~s$^{-1}$cm$^{-3}$Hz$^{-1}$sr$^{-1}$),
and the absorption coefficient, $\kappa_{\nu}$ (in cm$^{-1}$) are related by Kirchhoff's
law;   for combined modes of polarization, for unity spectral index and a 
frequency $\nu \ll 10^{10}T_{\rm eff}$ ($\nu$ in Hz, $T_{\mathrm{eff}}$ in K; 
Rayleigh-Jeans approximation):
\begin{equation}\label{kirchhoff}
{\eta_{\nu}\over \kappa_{\nu}} = {2kT_{\rm eff}\nu^2\over c^2}
                    \approx 3.1\times 10^{-37}T_{\rm eff}\nu^2.
\end{equation}
Here, $k=1.38\times 10^{-16}$~erg~K$^{-1}$ is the Boltzmann constant.
The brightness temperature $T_b$ is 
\begin{equation}
   T_b = \left\{  \begin{array}{ll}
                     \tau T_{\rm eff},   & \mbox{$\tau \ll 1$} \\
		          T_{\rm eff},   & \mbox{$\tau \gg 1$}
	            \end{array} 
         \right.    
\end{equation}
with $\tau = \int\kappa d\ell$ being the optical depth along the line of sight
$\ell$. The spectral radio flux density $S_{\nu}$ then is 
\begin{equation}\label{rayleighjeans}
S_{\nu} = {2kT_b\nu^2\over c^2}{A\over d^2} 
        \approx 0.1\left({T_b\over 10^6~\mathrm{K}}\right)
                   \left({\nu \over \mathrm{1~GHz}}\right)^2
                   \left({r\over 10^{11}~\mathrm{cm}}\right)^2
                   \left({1~\mathrm{pc}\over d}\right)^2~\mathrm{mJy}
\end{equation}
where  $A$ is the cross sectional source area perpendicular to the line 
of sight, and the approximation on the right-hand side is for a circular source with 
$A = \pi r^2$.
In non-thermal astrophysical plasmas, electron volumetric number densities are often 
observed to follow a power-law form
\begin{equation}\label{powerlaw}
n(\epsilon) = N(\delta-1)\epsilon_0^{\delta-1}\epsilon^{-\delta} 
\quad \mathrm{[cm^{-3}erg^{-1}]}
\end{equation}
where $\epsilon = (\gamma-1)m_ec^2$ is the kinetic particle energy,
$\gamma$ is the Lorentz factor,  and $\delta > 1$ has been assumed 
so that $N$ is the total non-thermal electron number density 
above $\epsilon_0$. A fundamental 
frequency of a plasma is its  plasma frequency, given by
\begin{equation}\label{plasmafrequency}
\nu_p \equiv {\omega_p\over 2\pi} = \left({n_ee^2\over \pi m_e}\right)^{1/2} 
\approx 9000~n_e^{1/2}\quad \mathrm{[Hz]}.
\end{equation}
Here, $m_{e} = 9.1\times 10^{-28}$~g is the electron rest mass, 
$e = 4.8\times 10^{-10}$~esu is the electron charge, and $n_e$ is
the total electron number density (cm$^{-3}$).

\subsection{Bremsstrahlung}

A thermal plasma  emits free-free emission (bremsstrahlung) across the electromagnetic 
spectrum. For cosmic abundances, the absorption coefficient is then
approximately given by  (in the usual units; $T \equiv T_{\mathrm{eff}}$, mix of 90\% H
and 10\% He; upper equation for singly ionized species, lower equation for fully ionized
plasma, and $\nu \gg \nu_p$;  \citealt{dulk85, lang99}):
\begin{equation}
\kappa_{\nu} \approx 0.01n_e^2T^{-3/2}\nu^{-2}\times
     \left\{  
        \begin{array}{ll}
              {\mathrm{ln}}(5\times 10^{7}T^{3/2}/\nu), &  
              \mbox{$T \la 3.2\times 10^5~\mathrm{K}$} \label{freefree_kappa2} \\ 
              {\mathrm{ln}}(4.7\times 10^{10}T/\nu), &
              \mbox{$T \ga 3.2\times 10^5~\mathrm{K}$} \label{freefree_kappa1}  
	\end{array} 
     \right.    
\end{equation}
and the emissivity $\eta_{\nu}$ follows from  equation~\ref{kirchhoff}.
Optically thick bremsstrahlung shows a $\nu^2$ dependence, whereas optically thin
flux is nearly independent of $\nu$. A homogeneous optically thin magnetized
source is polarized in the sense of the magnetoionic x-mode, whereas an optically 
thick source  is unpolarized \citep{dulk85}.
          
\subsection{Gyromagnetic Emission}

Electrons in magnetic fields radiate gyromagnetic emission.
The  gyrofrequency in a magnetic field of strength $B$ is
\begin{equation}\label{gyrofrequency}
\nu_c \equiv {\Omega_c\over 2\pi} = {eB\over 2\pi m_ec} \approx 
        2.8\times 10^6~B~\mathrm{[Hz]}
\end{equation}
where  $c = 3\times 10^{10}$~cm~s$^{-1}$ is the speed of light,
and the magnetic field strength $B$ is given in 
Gauss. The relativistic gyrofrequency is $\nu_{\mathrm{c, rel}} = 
\nu_{\mathrm{c}}/\gamma$. For large pitch angles, spectral power 
is predominantly emitted around a harmonic $s$ with
\begin{equation}\label{peaknu}
\nu_{\rm max} = s\nu_{c,\mathrm{rel}} \approx \gamma^3\nu_{c,\mathrm{rel}} 
       \approx 2.8\times 10^6B\gamma^2 \quad {\mathrm{[Hz]}}.
\end{equation}
Depending on $\gamma$, the emission is termed {\it gyroresonance} or {\it cyclotron}
emission ($s < 10$, $\gamma \approx 1$, non-relativistic, typically thermal 
electrons),  {\it gyrosynchrotron}   ($s \approx 10-100$, $\gamma \la 2-3$, mildly
relativistic electrons), or {\it synchrotron} emission ($s > 100$, 
$\gamma \gg 1$, relativistic electrons). Because relativistic effects
play an increasingly important role in increasing $s$, the
fundamental properties (e.g., directivity, bandwidth, polarization)
change between the three categories of gyromagnetic emission.
The total  power emitted by an electron is 
\citep{lang99}
\begin{equation}
P = 1.6\times 10^{-15}\beta^2\gamma^2B^2\mathrm{sin}^2\alpha
     ~\mathrm{[erg~s^{-1}]}
\end{equation}
where $\beta = v/c = (1-\gamma^{-2})^{1/2}$ and $\alpha$ is the pitch angle
of the electron.

Simplified approximate expressions for separate magnetoionic modes of
gyromagnetic emission have been
given by \citet{dulk82, robinson84}, and \citet{klein87a}. 
For stellar observations, the angle $\theta$ between the line of sight and the 
magnetic field is often unknown and should be averaged. We give  
handy, simplified expressions for $\eta_{\nu}$, $\kappa_{\nu}$, the 
turnover frequency $\nu_{\mathrm{peak}}$, and the degree of circular
polarization $r_c$ derived from \citet{dulk85} for $\theta = \pi/3$, with 
exponents for the model parameters $B, N, T,$  and the characteristic source scale
along the line of sight, $L$. For more comprehensive 
expressions, we refer to the original work by \citet{dulk82} and \citet{dulk85}. 

\medskip

\noindent {\it Gyroresonance emission:} If $s$ is the harmonic number,
            $s^2T \ll 6\times 10^9$~K and $\nu \ga \nu_c$, then for each 
	    magnetoionic mode (for unity refractive index),
\begin{eqnarray} 
\kappa_{\nu}(s) &=& 1020(1\pm 0.5)^2{n_e\over \nu T^{1/2}}{s^2\over s!}
                   \left({s^2T\over 1.6\times 10^{10}}\right)^{s-1}
                     \mathrm{exp}\left[-{ (1-s\nu_c/\nu)^2\over 
                     8.4\times 10^{-11}T}\right] \label{gyroresonance_kappa} \\
\eta_{\nu}(s) &=& {kT\nu^2\over c^2}\kappa_{\nu}(s)  \label{gyroresonance_eta}
\end{eqnarray}
(for the x-mode, and marginally for the o-mode, see \citealt{dulk85}) where $n_e$ is 
the ambient electron density. Properties: Emission is strongly concentrated in emission 
``lines'' at harmonic frequencies $s\nu_{\mathrm{c}}$, and $\kappa_{\nu}$
depends on the magnetoionic mode (lower sign for o-mode, upper sign for x-mode).
The emission at a given harmonic comes from a layer of constant $B$ with a thickness
$L = 2\Lambda_{\rm B}\beta_0\mathrm{cos}\theta \approx \Lambda_{\rm B}\beta_0$
determined by the magnetic scale length $\Lambda_{\rm B} = B/\nabla B$ and
$\beta_0 \equiv [kT/(m_ec^2)]^{1/2} \approx 1.3\times 10^{-5}T^{1/2}$.

\medskip
\newpage

\noindent {\it Gyrosynchrotron Emission from a Thermal Plasma:}
If $10^8 \la T \la 10^9$~K and $10 \la s \la 100$, then for the x-mode (and $r_c$
for  $\tau \ll 1$)

\begin{eqnarray}
\kappa_{\nu} &\approx& 21n_eT^7B^9\nu^{-10}                \label{thermalsynchrotron_kappa}\\
\eta_{\nu} &\approx& 3.2\times 10^{-36}n_eT^8B^9\nu^{-8}   \label{thermalsynchrotron_eta}\\
\nu_{\rm peak} &\approx& 1.3(n_eL/B)^{0.1}T^{0.7}B         \label{thermalsynchrotron_peak}\\
 r_c  &\approx& 2.9\times 10^4T^{-0.138}(\nu/B)^{-0.51}.   \label{thermalsynchrotron_pol}
\end{eqnarray}
Properties: The spectral power is $\propto \nu^2$ on the optically thick side, 
but $\propto \nu^{-8}$ on the optically thin side of the spectrum for a homogeneous source.

\medskip

\noindent {\it Gyrosynchrotron Emission from a Power-Law Electron Distribution:}
For isotropic pitch angle electron distributions according to equation~\ref{powerlaw}, 
with $\epsilon_0 = 10~\mathrm{keV} = 1.6\times 10^{-8}$~erg,  $2 \la \delta  \la 7$, 
and $10 \la s \la 100$, and for the x-mode (and $r_c$ for  $\tau \ll 1$)

\begin{eqnarray} 
\kappa_{\nu}   &\approx& 10^{-0.47+6.06\delta}NB^{0.3+0.98\delta}\nu^{-1.3-0.98\delta}  \label{gyrosynchrotron_kappa}\\
\eta_{\nu}     &\approx& 10^{-31.32+5.24\delta}NB^{-0.22+0.9\delta}\nu^{1.22-0.9\delta} \label{gyrosynchrotron_eta}\\
\nu_{\rm peak} &\approx& 10^{3.41+0.27\delta}(NL)^{0.32-0.03\delta}B^{0.68+0.03\delta}  \label{gyrosynchrotron_peak}\\
r_c            &\approx& 10^{3.35+0.035\delta}(\nu/B)^{-0.51}.                          \label{gyrosynchrotron_pol} 
\end{eqnarray}
Properties: Broad spectra with intermediate circular polarization degree on the  optically 
thin side. Optically thick spectral power is approximately $\propto \nu^{5/2}$, and the optically 
thin side is a power  law with an index $\alpha = 1.22-0.9\delta$  for a homogeneous source.

\medskip

\noindent {\it Synchrotron Emission from a Power-Law Electron Distribution} (homogeneous and
isotropic): For $\gamma \gg 1$, i.e., $s \gg 100$, in each of the magnetoionic modes
\begin{eqnarray}
\kappa_{\nu} &\approx& 10^{5.89+1.72\delta}(\delta-1)NB^{(\delta+2)/2}\nu^{-(\delta+4)/2} \label{synchrotron_kappa} \\
\eta_{\nu} &\approx& 10^{-24.7+1.57\delta}(\delta-1)NB^{(\delta+1)/2}\nu^{-(\delta-1)/2} \label{synchrotron_eta} \\
\nu_{\rm peak} &\approx& \left[10^{11.77+3.44\delta}(\delta-1)^2N^2L^2B^{\delta+2}\right]^{1/(\delta+4)} \label{synchrotron_peak}
\end{eqnarray}
Properties: Continuous and broad spectrum, important harmonics $s \approx 
\gamma^3$. For a homogeneous, optically thin source, the degree of  linear
polarization is $r = (\delta+1)/(\delta + 7/3)$. The optically thick (self-absorbed) spectral 
power is $\propto \nu^{5/2}$, and the optically thin side of the spectrum has a power-law
spectral index of   $\alpha = -(\delta - 1)/2$.

\subsection{Coherent Emission}\label{coherent}

The brightness temperature of synchrotron emission is limited to $T_b \la 10^{12}$~K by
inverse Compton scattering \citep{kellermann69}.
If higher $T_b$ is observed, then a  coherent radiation
mechanism should be considered. Two  mechanisms have received most attention 
both for solar and stellar coherent bursts (for details, see \citealt{benz02}): 

Plasma radiation is emitted at the fundamental or the second harmonic 
of  $\nu_p$ (equation~\ref{plasmafrequency}; for propagation parallel
to the magnetic field), or of the upper hybrid frequency $\nu_{\mathrm{UH}} = 
(\nu_p^2 + \nu_c^2)^{1/2}$ (for perpendicular propagation, with $\nu_p > \nu_c$). 
It is thus a useful means to approximately determine the 
electron density in the source.
It can account for high brightness temperatures (up to $10^{18}$~K, 
\citealt{melrose89}) and small bandwidth, and it is frequently observed in the Sun
\citep{bastian98} and in low-frequency stellar flares \citep{bastian90}. Owing to 
increasing  free-free absorption with increasing $\nu$
(because $n_e \propto \nu_p^2$ in equation~\ref{freefree_kappa1}), and possibly
also owing to gyroresonance absorption in hot plasma, fundamental emission  
is best observed below 1~GHz, although in very-high temperature
environments such as coronae of RS CVn binaries, the absorption is
milder and the limitations are more relaxed \citep{white95}.
The escape of fundamental emission is also alleviated in a highly structured medium 
with sharp gradients in density \citep{aschwanden87, benz92}. 

Electron cyclotron maser emission  \citep{melrose82, melrose84} is emitted mostly 
at the fundamental  and the second harmonic of $\nu_c$ 
(equation~\ref{gyrofrequency}) and can therefore be used to determine the 
magnetic field strength in the source. The requirement for radiation 
propagation, $s\nu_c > \nu_p$, also sets an upper limit to the electron density 
in the source and along the line of sight to the observer.
The cyclotron maser mechanism  accounts for the observed high $T_b$
($\la 10^{20}$~K for $s=1, \la 10^{16-17}$~K for $s=2$), and polarization 
degrees up to $100\%$.

Other mechanisms have occasionally been proposed. For example, plasma maser emission
from a collection of double layers was initially calculated by \citet{kuijpers89b} 
and applied to highly polarized stellar flare emission by \citet{vdoord94}.

\subsection{Wind emission}

\citet{olnon75, panagia75}, and \citet{wright75} calculated the radio spectrum of 
a spherically symmetric, ionized wind. If $\dot{M}$ is the 
mass loss rate in units of $M_{\odot}$yr$^{-1}$, $v$  the terminal wind velocity 
in km~s$^{-1}$, $d_{\mathrm{pc}}$  the distance to the star in pc, and $T$ the wind 
temperature in K, the optically thick radius is (we assume a mean atomic weight per 
electron of 1.2, and an average ionic charge of 1)
\begin{equation}\label{windsurface}
R_{\mathrm{thick}} = 8\times 10^{28}(\dot{M}/v)^{2/3}T^{-0.45}\nu^{-0.7}\quad\mathrm{[cm]}.
\end{equation}
Then, the following formulae apply ($R_*$ in cm):
  \begin{equation}
    S_{\nu} = \left\{  \begin{array}{ll}
        9\times 10^{10}(\dot{M}/v)^{4/3}T^{0.1}d_{\mathrm{pc}}^{-2}\nu^{0.6}
	~\mathrm{[mJy]},   
        & \mbox{if $R_{\mathrm{thick}} \ge R_*$}  \\
	5\times 10^{39}(\dot{M}/v)^{2}T^{-0.35}R_*^{-1}d_{\mathrm{pc}}^{-2}
	\nu^{-0.1}~\mathrm{[mJy]},   
	& \mbox{if $R_{\mathrm{thick}} < R_*$.}
        \end{array} 
        \right. \label{wind}   
  \end{equation}

\subsection{Loss Times}

Relativistic electrons in an ambient gas of density $n_e$ lose energy by Coulomb collisions
with the ions. The energy loss rate and the corresponding life time are
(under typical coronal conditions; \citealt{petrosian85})
\begin{equation}\label{collloss}
-{\dot{\gamma}}_{\mathrm{coll}} 
     = 5\times 10^{-13}n_e\quad \mathrm{[s^{-1}]}, \quad\quad \tau_{\mathrm{coll}}   
     = 2\times 10^{12}{\gamma\over n_e} \quad \mathrm{[s]}.
\end{equation}
For electrons in magnetic fields, the synchrotron
loss rate and the life time are given by (\citealt{petrosian85}, pitch angle = $\pi/3$)
\begin{equation}\label{synchloss}
-\dot{{\gamma}}_{\mathrm{B}}  
     = 1.5\times 10^{-9}B^2\gamma^2\quad \mathrm{[s^{-1}]},\quad\quad \tau_{\mathrm{B}} = 
     {6.7\times 10^8\over B^2\gamma}\quad \mathrm{[s]}.
\end{equation}

\section{RADIO FLARES FROM COOL STARS}\label{flares}

\subsection{Incoherent Flares}

Many radio stars have been found to be flaring sources,
and a considerable bibliography on stellar radio flares is now available.
As in the Sun, two principal flavors are present: incoherent
and coherent radio flares.
Incoherent flares with time scales of minutes to hours,
broad-band spectra, and moderate degrees of polarization  are thought to be 
the stellar equivalents of solar microwave bursts.
Like the latter, they show evidence for the presence of mildly 
relativistic electrons radiating gyrosynchrotron emission in magnetic fields.  
Many flares on single F/G/K stars are of this type \citep{vilhu93, guedel95b, guedel98}, 
as are almost all radio flares on M dwarfs above 5~GHz \citep{bastian90, guedel96a},
on RS CVn binaries \citep{feldman78, mutel85b}, on contact binaries
\citep{vilhu88}, and on  other active subgiants and giants 
although some of the latter perhaps stretch the ``solar analogy'' 
 beyond the acceptable limit: The FK Com-type giant HD~32918
produced flare episodes lasting  2$-$3 weeks, with a radio luminosity
of $6\times 10^{19}$~erg~s$^{-1}$Hz$^{-1}$ \citep{slee87b, bunton89}. Its integrated
microwave luminosity is thus about 1000 times larger than the total  X-ray output
of the non-flaring Sun!  
      
\subsection{Coherent Radio Bursts}

Bursts that exhibit the typical characteristics of coherent emission
(see ``Coherent Emission,'' above) probably represent stellar equivalents of metric and 
decimetric solar bursts that themselves come in a complex variety
\citep{bastian98}. Some exceptionally long-lasting ($\sim$1~h) but highly 
polarized  bursts require a coherent mechanism as well
(examples in \citealt{lang86a, white86, kundu87, willson88}, and \citealt{vdoord94}). 
Coherent flares are abundant on M dwarfs at longer wavelengths (20~cm; 
\citealt{kundu88, jackson89}). There are  also some interesting
reports on highly polarized, possibly coherent bursts in RS CVn binaries 
\citep{mutel78, brown78, simon85, lestrade88, white95, jones96, dempsey01}.
After many early reports on giant metric flares observed with
single-dish telescopes \citep{bastian90}, 
the metric wavelength range has subsequently been largely neglected, with
a few notable exceptions \citep{kundushev88, vdoord94}.

Coherent bursts carry profound information in high-time resolution light curves. 
Radio ``spike'' rise times as short as 5$-$20~ms have been reported, implying
source sizes of $r< c\Delta t \approx  1500$-$6000$~km 
and brightness temperatures up to $T_b \approx 10^{16}$~K, a clear proof of the 
presence of a coherent mechanism \citep{lang83, lang86b, guedel89a, bastianea90}.
Quasi-periodic oscillations were found
with time scales of 32~ms to 56~s \citep{gary82, lang86b, bastianea90}, and up to
$\approx$5~min  during a very strong flare  (\citealt{brown78}, although 
    this emission was proposed to be gyrosynchrotron radiation).

\subsection{Radio Dynamic Spectra}

\begin{figure}[t!] 
\centerline{\psfig{file=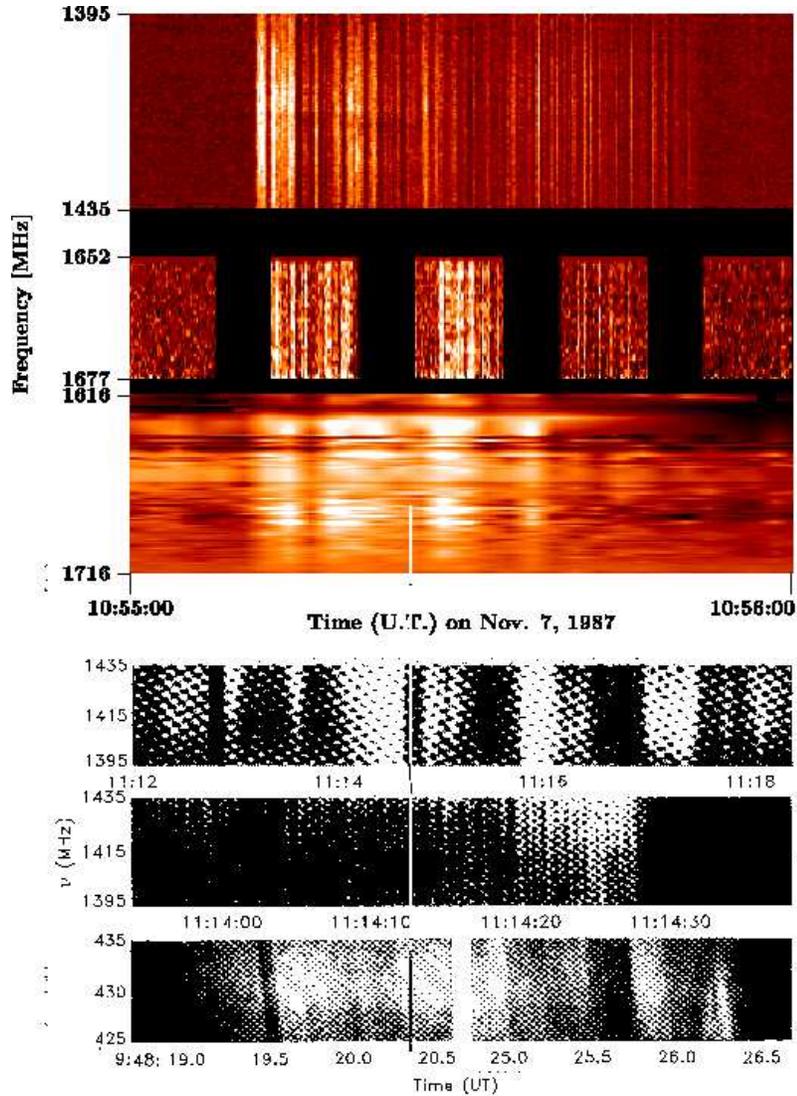,width=10.7truecm}}
\vspace*{-5mm} \caption{Gallery of radio dynamic spectra of M dwarf flares. Upper three panels show
a  flare on  AD Leo, recorded with the Arecibo (top), Effelsberg (middle) and 
Jodrell Bank (bottom) telescopes in different wavelength ranges
(see also \citealt{guedel89a}). Bottom three panels show flares on
AD Leo (top and middle) and YZ CMi (bottom) observed at Arecibo
(after \citealt{bastianea90}, reproduced with permission of the AAS).\label{dyn}}
\end{figure}

If the elementary frequencies relevant for
coherent processes, $\nu_p$ and $\nu_c$, evolve in the source, or the radiating 
source itself  travels across density or magnetic field gradients, the emission  
leaves characteristic traces on the $\nu-t$ plane, i.e., on  dynamic spectra
\citep{bastian98}. A rich phenomenology  has been uncovered, including: a)
short, highly polarized bursts with structures as narrow as $\Delta\nu/\nu = 0.2\%$
suggesting either plasma emission from a source of size $\sim 3\times 10^8$~cm, or a cyclotron 
maser in magnetic fields of $\sim 250$~G \citep{bastian87, guedel89a, bastianea90}.
b) Evidence for spectral structure with positive
drift rates of 2 MHz~s$^{-1}$ around 20~cm wavelength, taken as evidence
for a disturbance propagating ``downward'' in the plasma emission interpretation
\citep{jackson87}; and c) in solar terminology,
rapid broadband pulsations, ``sudden (flux) reductions'', and 
positive and negative drift rates of 250$-$1000~MHz~s$^{-1}$
(\citealt{bastianea90, abada94, abada97}, see Figure~\ref{dyn}).  

The smallest spectral bandwidths were found to be in the 1\% range for some bursts.  
Conservatively assuming a magnetic scale height of $\Lambda_B = 1R_*$, the source size 
can be estimated to be $r \approx (\Delta\nu/\nu)\Lambda_B \approx $ a few 1000~km, 
again implying very high $T_b$ \citep{lang88a, guedel89a, bastianea90}.

\section{QUIESCENT EMISSION FROM CORONAE OF COOL STARS}

\subsection{Phenomenology}\label{quiescent}

The discovery of quiescent radio emission from magnetically active stars 
was one of the most fundamental and one of the least anticipated achievements 
in radio astronomy of cool stars because there simply is no solar counterpart. 
The Sun emits steady, full-disk, optically thick thermal radio emission at chromospheric and
transition region levels of a few $10^4$~K. However, one derives from 
equation~\ref{rayleighjeans} that such emission cannot be detected with present-day facilities, 
except for radiation from the very nearest stars, or giants subtending 
a large solid angle (see ``Radio Emission from Chromospheres and Winds,'' below). 
The radio luminosity of the corona caused by optically thin bremsstrahlung
is proportional to the observed X-ray emission measure, but again 
the calculated radio fluxes are orders of magnitude below observed fluxes
\citep{gary81, topka82, kuijpers85, borghi85, massi88, guedel89b, vdoord89}.
Obviously, another mechanism is in charge, and its characterization requires
a proper description of the phenomenology.

Quiescent emission  is best defined by the absence of impulsive,
rapidly variable flare-like events. Common characteristics of quiescent 
emission are (i) slow variations on time scales of hours and days 
\citep{pallavicini85, lang86a, lang88a, willson88, jackson89}, (ii) broad-band 
microwave spectra \citep{mutel87, guedel89b},  (iii) brightness temperatures 
in excess of coronal temperatures measured in X-rays \citep{mutel85b, lang88a, white89a}, 
and often (iv) low polarization degrees \citep{kundu88, jackson89, drake92}. 
Occasionally,  strong steady polarization up to 50\% \citep{linsky83, pallavicini85, 
willson88, jackson89} or unexpectedly narrow-band steady emission \citep{lang86a, lang88a}
is seen on M dwarfs.
Quiescent emission has been reported at frequencies as low as 843~MHz 
\citep{vaughan86, large89} and as high as 40~GHz \citep{white90}.

\subsection{Gyromagnetic Emission Mechanisms}\label{emissionmechanisms}

\begin{figure}[t!]
\centerline{\psfig{file=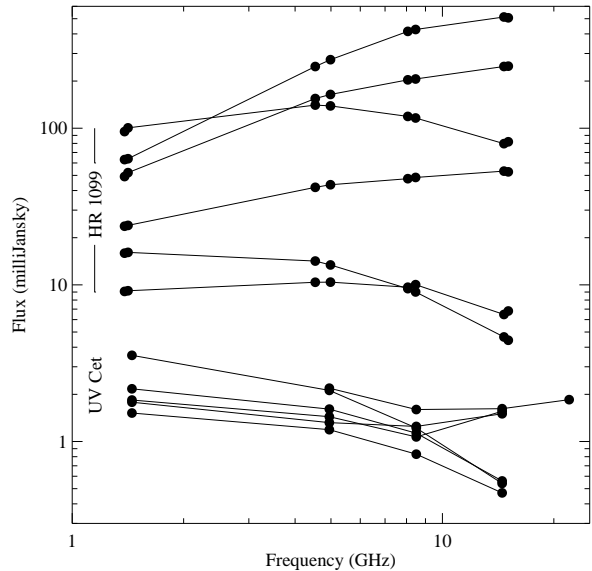,width=8truecm}}
\caption{Radio spectra of the RS CVn binary HR~1099  (upper set) and of the dMe dwarf  
UV Cet (lower set) at different flux levels. The gently bent spectra are indicative
of gyrosynchrotron emission, and the high-frequency part of U-shaped spectra for 
UV Cet has been interpreted as a gyroresonance component (HR~1099 spectra: courtesy
of SM White).\label{spec}}
\end{figure}

Because active stars show high coronal temperatures and large magnetic filling factors 
\citep{saar90} that prevent magnetic fields from strongly diverging with increasing height, 
the radio optical depth can become significant at coronal levels
owing to  gyroresonance absorption (equation~\ref{gyroresonance_kappa}). Such emission
is also observed above solar sunspots. The emission occurs at low harmonics of $\nu_c$, 
typically at $s=3-5$ \citep{guedel89b, vdoord89, white94}.
Rising spectra between 5 and 22~GHz occasionally observed on stars support such an 
interpretation  \citep{cox85} and imply the presence of $\sim 10^7$~K plasma in 
strong, kG magnetic fields (\citealt{guedel89b}; Figure~\ref{spec}).  However, in most cases
high-frequency  fluxes are  lower than predicted from  a uniform,
full-disk gyroresonance layer, constraining the filling factor of strong magnetic fields 
immersed in hot plasma. It is possible that the hot plasma resides in low-$B$ 
field regions where the gyroresonance absorption is negligible, for example
between strong fields from underlying active regions, or at large heights
\citep{white94, lim96a}. Uniform magnetic structures of $\sim$600~G containing hot 
plasma can, on average, not reach out to beyond 1$-$2$R_*$ on M dwarfs \citep{leto00}.

The gyroresonance  mechanism cannot  apply to 
lower-frequency radio emission, because the radius of the optically thick layer, still
at $s = 3-5$, would have to be  $ > 3R_*$  for dMe stars. However, an extrapolation of 
the corresponding magnetic fields of  more than 100~G down to photospheric levels 
would result in photospheric fields much stronger than observed 
\citep{gary81, topka82, linsky83, pallavicini85, lang86a, willson88, guedel89b}. 

The only remedy is to  allow for much higher $T_{\mathrm{eff}}$. Then,  the optically 
thick layer shifts to harmonics above 10, the range of gyrosynchrotron radiation,
and the optically thick source sizes become more reasonable for M dwarfs, 
$R \approx R_*$ \citep{linsky83, drake89, drake92}. For a thermal plasma, 
however, the spectral power drops like $\nu^{-8}$ at high 
frequencies (equation~\ref{thermalsynchrotron_eta}), 
far from the rather shallow $\nu^{-(0.3...1)}$ spectra  of active stars 
(\citealt{massi92}; Figure~\ref{spec}). On the other hand, an optically thick thermal
contribution  requires strong magnetic fields ($\sim$200~G) over such large source areas 
that inferred photospheric magnetic fields again become unrealistically large 
\citep{kuijpers85, drake92}.

The situation is much more favorable for a  non-thermal electron  
energy distribution such as  a power law, analogous to distributions inferred
for solar and stellar microwave flares \citep{kundu85, pallavicini85}. This
model is supported by measured high brightness temperatures 
\citep{lestrade84b, mutel85b, umana91, benz95}. Comprehensive 
spectral modeling suggests mildly relativistic electrons in $\sim$100~G fields
with power-law indices of $\delta \approx 2-4$, matching observed broad spectra with turnover 
frequencies in the $1-10$~GHz range (\citealt{white89b, slee87a, slee88, umana98}; 
Figure~\ref{spec}). The  non-thermal model is now quite well established for
many classes of active stars and provokes the question of how these coronae are 
continuously replenished with high-energy electrons.

Occasionally,  steady narrow-band or strongly polarized coherent emission 
has been observed at low frequencies. An interesting possibility is large numbers 
of unresolved, superimposed coherent bursts \citep{pallavicini85, lang86a, lang88a, 
white95, large89}.

\section{RADIO FLARES AND CORONAL HEATING}\label{coronalheating}

\subsection{Is Quiescent Emission Composed of Flares?}\label{longterm}

The question of the nature of stellar quiescent radio emission has defined one of
the most fascinating aspects of stellar radio astronomy. A number of observations
seem to suggest that flares play an important role - quiescent emission
could simply be made up of unresolved flares. Inevitably, this question relates to the physics of
coronal heating and the presence of X-ray coronae. 

Very Long Baseline Interferometry (VLBI)  studies of RS CVn binaries suggest that 
``flare cores'' progressively expand into 
a large-scale magnetosphere around the star, radiating for several days (``flare remnant emission'').
The electron distributions then evolve from initial power-law distributions
as they are subject to collisional losses affecting the low-energy electrons and  to 
synchrotron losses affecting the high-energy electrons. Time-dependent calculations
predict rather flat spectra similar to those observed \citep{chiuderidrago93, franciosini95,
torricelli98}.

\subsection{Microflaring at Radio Wavelengths}

\begin{figure}[t!]
\centerline{\psfig{file=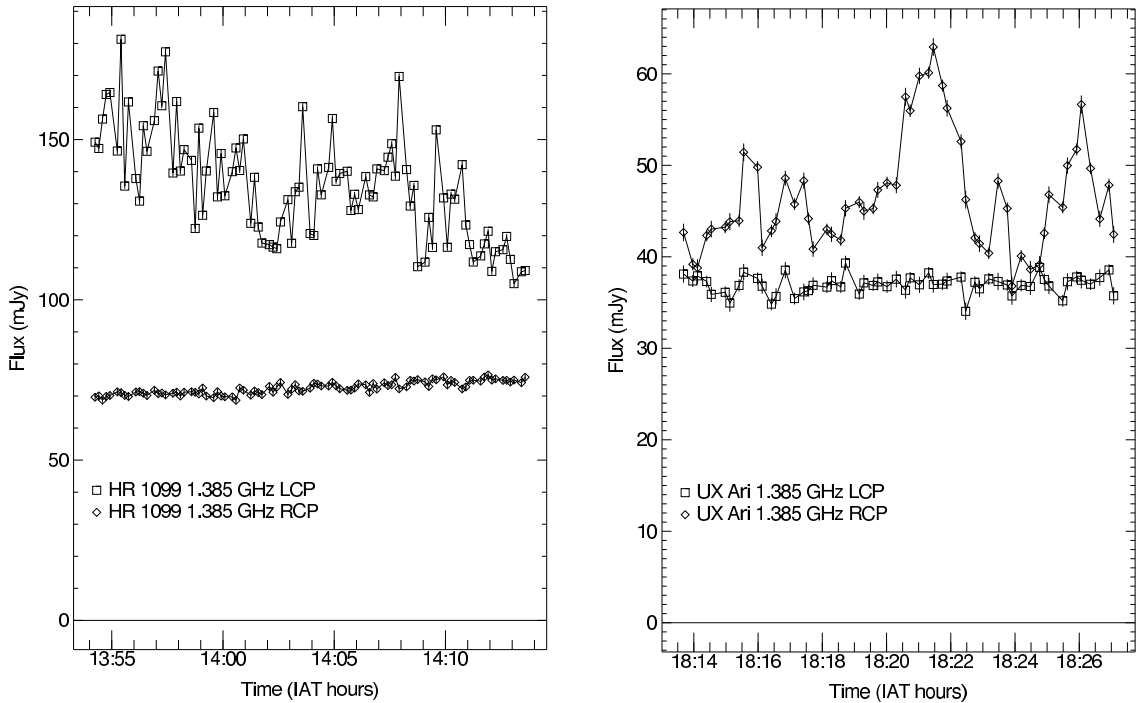,width=15truecm}}
\caption{Light curves of HR~1099 ({\it left}) and UX Ari ({\it right}) obtained
in the two senses of circular polarization. The brighter of the two polarized
fluxes varies rapidly and has been interpreted as 100\% polarized 
coherent emission superimposed on a gradually changing gyrosynchrotron component 
(White \& Franciosini 1995; figures from SM White).\label{light}}
\end{figure}

Very long quiescent episodes with little flux changes impose challenges for flare-decay 
models. Limitations on the collisional losses (equation~\ref{collloss}) 
require a very low ambient electron density \citep{massi92}. Frequent electron injection 
at many sites may be an alternative. Based on spectral observations, \citet{white95} 
suggest that the emission around 1.4~GHz is composed of a steady, weakly polarized 
broad-band gyrosynchrotron component plus superimposed, strongly and oppositely 
polarized, fluctuating plasma emission that is perceived as quasi-steady but that may
occasionally evolve into strong, polarized flare emission 
(\citealt{simon85, lestrade88, jones96, dempsey01}; Figure~\ref{light}).

RS CVn and Algol-type binaries also reveal  significant, continuous gyrosynchrotron variability 
on time scales of $\sim$10$-$90 minutes during $\sim$30\% of the time, with an
increasing number of events with decreasing flux \citep{lefevre94}. These flare-like events
may constitute a large part of the quiescent gyrosynchrotron emission. However, ``microbursts''
 or ``nanoflares'' with durations on the order of seconds to a few minutes are usually
not detected  at available sensitivities  \citep{kundu88, rucinski93}
although \citet{willson87} found some rapid variability on time scales between 30~s and 1~h
that they interpreted as being due to variable absorption by thermal plasma.

\subsection{Radio Flares and the Neupert Effect}\label{neuperteffect}

\begin{figure}[t!]
\hbox{
\hskip 1truecm\psfig{file=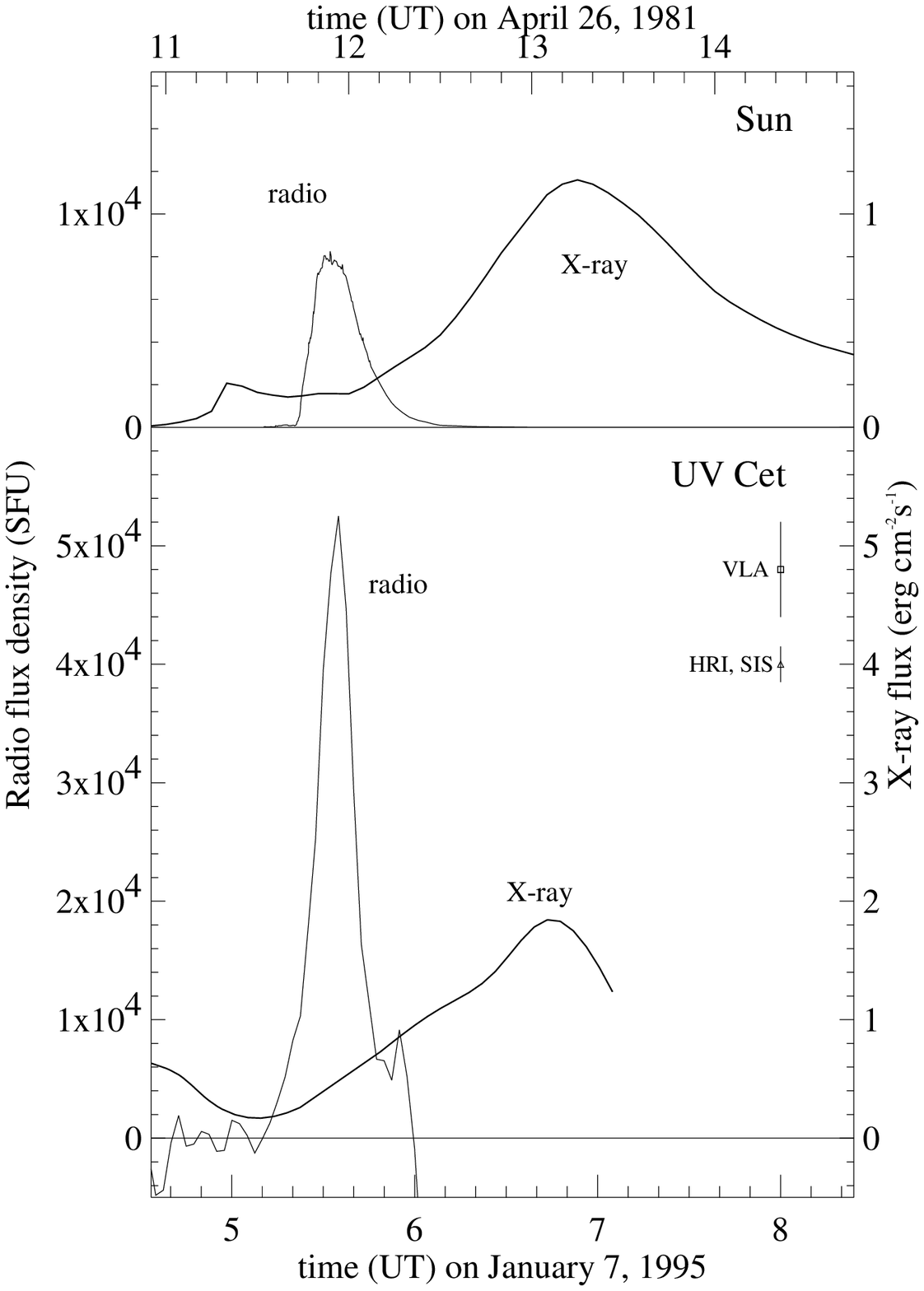,width=6.97truecm}
\hskip 1truecm\psfig{file=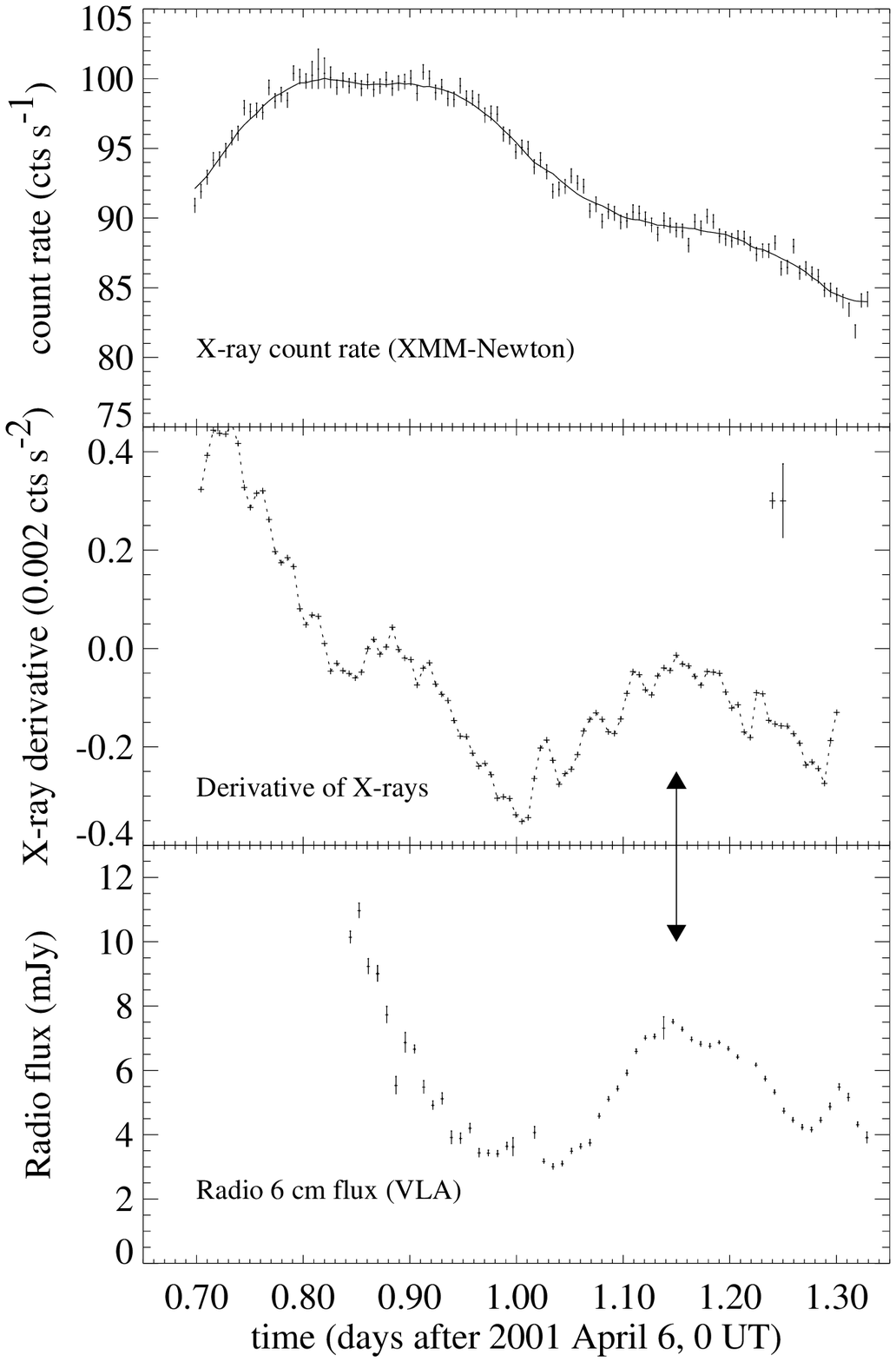,width=6.57truecm}
}
\caption{{\it Left}: Neupert effect seen in an M dwarf star, compared with a solar
example in the upper panel \citep{guedel96a}. {\it Right}: Neupert effect seen
in an RS CVn binary during a large flare \citep{guedel02}.\label{neupertfig}}
\end{figure}

The chromospheric evaporation scenario devised for many solar flares assumes
that accelerated coronal electrons precipitate
into the chromosphere where they lose their kinetic energy by collisions, thereby
heating the cool plasma to coronal flare temperatures and evaporating it
into the corona. The radio gyrosynchrotron emission from the accelerated electrons is roughly 
proportional to the injection rate of electrons, whereas the X-ray luminosity is roughly 
proportional to the accumulated energy in the hot plasma. To first order, one expects
\begin{equation}\label{neupert}
L_R(t) \propto {d\over dt}L_{\rm X}(t),
\end{equation}
a relation that is known as the ``Neupert Effect'' \citep{neupert68} and that has been well
observed on the Sun in most impulsive and many gradual flares \citep{dennis93}. The search
for stellar equivalents has been a story of contradictions if not desperation (Figure~\ref{neupertfig}).
A first breakthrough came with  simultaneous EUV (a proxy for X-rays) and optical 
observations (a proxy for the radio emission) of a flare on AD Leo \citep{hawley95} 
and  radio + X-ray observations of flares on UV Cet (\citealt{guedel96a}; 
Figure~\ref{neupertfig}a). 
The latter revealed a relative timing between the emissions that is very similar to solar 
flares. Also,  the energy ratios seen in 
non-thermal and thermal emissions are similar to solar analogs but, perhaps more interesting,
they are also similar to the corresponding ratio for the quiescent losses. These observations 
suggest  that high-energy particles are  the ultimate cause for heating through chromospheric 
evaporation not only in flares, but also in the ``quiescent'' state.  In 
retrospect, further suggestive examples of the Neupert effect can be found in the published
literature, most notably in \citet{vilhu88, stern92, brown98}, or \citet{ayres01}.

It is important to note that the Neupert effect is observed neither in each solar 
flare (50\% of solar  gradual flares show a different behavior; Dennis \& Zarro 1993),
nor in each stellar flare. Stellar counter-examples include an impulsive optical flare 
with following gradual radio emission \citep{vdoord96b}, gyrosynchrotron emission that 
peaks after the soft X-rays \citep{osten00}, and an X-ray depression during strong radio flaring
\citep{guedel98}. Note also that complete absence of correlated flaring has been observed 
at radio and UV wavelengths (e.g., \citealt{lang88b}).

\subsection{The Correlation between Quiescent Radio and X-Ray Emissions}

\begin{figure}[t!]
\centerline{\psfig{file=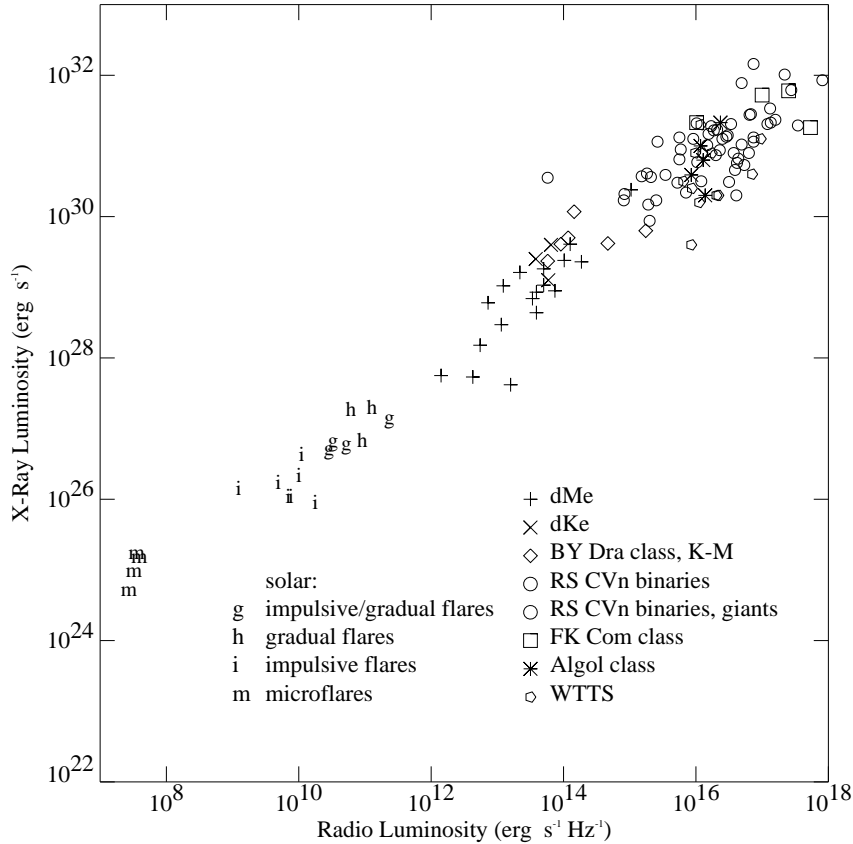,width=11.5truecm}}
\caption{Correlation between quiescent radio and X-ray luminosities of magnetically
active stars (symbols) and solar flares (letters; after \citealt{benz94}).\label{lxlr}}
\end{figure}

Radio detections are preferentially made among the X-ray brightest stars 
\citep{white89a}, a result that is corroborated by new, unbiased all-sky surveys 
\citep{helfand99}. For RS CVn and Algol binaries, one finds a correlation between the 
radio and X-ray luminosities, $L_{\mathrm{R}} \propto L_{\mathrm{X}}^{1.0-1.3}$
\citep{drake86a, drake89, drake92, fox94, umana98}, whereas for late-type dwarfs,
a rather tight linear correlation appears to apply
\citep{guedel93b}. In fact, several classes of active stars  
and solar flares follow a similar relation. Overall, for 5$-$8~GHz emission, 
\begin{equation}\label{lxlrrelation}
L_{\rm X}/L_{\rm R} \approx 10^{15\pm 1}\quad[\mathrm{Hz}]
\end{equation}
(\citealt{guedel93a, benz94}; Figure~\ref{lxlr}). Phenomenologically, the relation simply suggests 
that both radio and X-ray emissions are activity indicators that  reflect the level 
of magnetic activity. The most straightforward physical model, proposed 
by \citet{drake89, drake92}, assumes that hot plasma emits both thermal X-rays and 
non-thermal gyrosynchrotron radiation, but this model predicts
steep ($\propto \nu^{-8}$) optically thin spectra that are not observed.

\citet{chiuderidrago93} argued that the build-up of magnetic
energy in magnetic loops is $\propto B^2$ and proportional to the gradient
of the turbulent velocity (assumed to be similar in all stars).
Since most of the energy in the magnetic fields  is eventually radiated away
in X-rays, $L_{\rm X}\propto B^2V$, where $V$ is the source volume. A
power-law electron population in the same loop emits optically thin radio radiation as 
$L_{\mathrm{R}} \propto B^{-0.22 + 0.9\delta}V$ for given $\nu$
(equation~\ref{gyrosynchrotron_eta}). 
Therefore, $L_{\mathrm{R}} \propto L_{\mathrm{X}}^{0.45\delta - 0.11}V^{1.11-0.45\delta}$.
For $\delta = 3$, the dependence on $V$ is small and $L_{\mathrm{R}} \propto L_{\rm X}^{1.24}$, 
close to the reported relations. 

If the quiescent emission is made up of numerous solar-like coronal flares,
then the high-energy particles carry their
energy downward to heat chromospheric plasma.  For a steady-state
situation in which the electron injection rate is  balanced by an unspecified 
electron loss mechanism, one derives a relation between the 
synchrotron losses, the particle life time, and the energy losses in X-rays. 
Using the empirical correlation, \citet{guedel93a} estimated an 
electron life time that indeed corresponds quite well to the 
observed radio variability time scales. \citet{holman85, holman86}, and \citet{airapetian98} 
presented detailed model calculations for the alternative situation of 
simultaneous heating and particle acceleration in current sheets.

\section{STELLAR MAGNETIC FIELDS}\label{bfields}

Stellar radio astronomy has provided invaluable information on stellar coronal
magnetic fields not accessible by any other methods. We briefly describe
a few principal results. Electron cyclotron maser emission can be used to estimate the
field strength in the source region, namely in converging, lower-coronal fields. Because 
the radiation is emitted at the fundamental ($s=1$) or the second harmonic ($s=2$) of the 
gyrofrequency, $B = \nu_{\mathrm{Hz}}/(s\times 2.8\times 10^6) \approx 250-500$~G 
\citep{guedel89a, bastianea90}.

Coronal gyroresonance emission detected at 8$-$23~GHz is optically  thick at harmonics $s = 3-5$ 
of $\nu_B$ (see ``Gyromagnetic Emission Mechanisms,'' above), hence 
$B = \nu_{\mathrm{Hz}}/(s\times 2.8\times 10^6) 
\approx 600-2070$~G in low-lying loops in active regions \citep{guedel89b}. 

Large coronal sizes as seen by VLBI restrict
field strengths to typically 10$-$200~G if the  approximate dipolar extrapolation to the photosphere
should not  produce excessive surface field strengths  \citep{benz98, mutel98, beasley00}. 
     
Very high synchrotron brightness temperatures $T_b$ constrain, together with
the power-law index $\delta$, the characteristic harmonic numbers $s$ and therefore
the field strength \citep{dulk82}. 
\citet{lestrade84a, lestrade88,  mutel84, mutel85b}, and \citet{slee86}
found $B$ between  $\approx 5$~G and  several tens of G for RS CVn halo
sources, and a few tens to several hundred G  for the core.     
Further, the electron energy is $\epsilon = kT_b$ for optically thick emission, and 
$\epsilon > kT_b$ for optically thin emission.  Equation~\ref{peaknu} for
synchrotron emission ($s \gg 1$) then implies
   \begin{equation}\label{bfromt}
    B \le {2\pi m_e^3c^5\nu\over e (m_ec^2 + kT_b)^2} 
    \approx 3.6\times 10^{-7} 
                 {\nu\over (1 + 1.7\times 10^{-10}T_b)^2}~\mathrm{[G]}
   \end{equation}
where the $<$ sign applies for optically thin emission. Alternatively, from 
\citet{dulk82} one finds for gyrosynchrotron emission
$T_b \le T_{\mathrm{eff}} = 10^{6.15-0.85\delta}(\nu/B)^{0.5+0.085\delta}$.
Such arguments lead to field strengths of a few tens to a few hundred G in RS CVn binaries
\citep{mutel85b, lestrade88}.
 
The synchrotron turnover frequency $\nu_{\mathrm{peak}}$ determines $B$ through  
$B = 2.9\times 10^{13} \nu_{\mathrm{peak}}^5\theta^4S^{-2}$ where $\theta$ 
is the source diameter in arcsec, $S$ is the radio flux density in mJy 
at $\nu_{\mathrm{peak}}$ which is given in units of GHz here (\citealt{lang99}, after 
\citealt{slish63}).  For a typical RS CVn or dMe coronal source with a size of 1 
milliarcsec (mas)
as measured by VLBI, a turnover at $\nu_{\mathrm{peak}} \approx 5$~GHz
and a flux of 10~mJy, magnetic fields of up to a few hundred G are inferred.  
For  dipolar active region gyrosynchrotron models, \citet{white89b}
derived $B \approx 150\nu_{\mathrm{peak}}^{1.3}$ ($\nu_{\mathrm{peak}}$ in GHz),
which again implies field strengths of several hundred G for turnover frequencies in the GHz 
range.

Full spectral modeling of the magnetic field strength $B$, the non-thermal particle 
density, the geometric size, and the electron power-law index $\delta$
can constrain some of these parameters. \citet{umana93, umana99} and
\citet{mutel98} found $B \approx 10-200$~G in  their models for core plus halo 
structures in Algol-type stars.

\section{RADIO CORONAL STRUCTURE}

\subsection{Introduction}

 Although the photospheric
filling factor of kG magnetic fields on  the Sun is small, it can exceed 50\% on
late-type active stars. On the Sun,  magnetic fields can
rapidly expand above the transition region and thus drop below kG levels. On 
magnetically active stars, such divergence is prevented
by adjacent magnetic flux lines so that strong magnetic fields may exist also at coronal
levels. The effective scale height of the coronal  magnetic field however
also largely depends on the structure of the magnetic flux lines, i.e., on
whether they are compact loops with short baselines, long loops connecting
distant active regions, or large dipolar magnetospheres anchored at the 
stellar poles. This issue is unresolved. Arguments for small-scale coronal active
regions as well as for star-sized global magnetospheres have been put forward
(see discussions in \citealt{white89b, morris90}, and \citealt{storey96}). 

\subsection{Radio Eclipses and Rotational Modulation}\label{eclipses}

Eclipses and rotational modulation offer reliable information on radio
source geometries, but neither are frequently seen.
The radio sources may be  much larger 
than the eclipsing star, or  they may not be within the eclipse zone.
Complete absence of radio eclipses has, for example, been reported for 
AR Lac \citep{doiron84},  Algol \citep{vdoord89, mutel98}, and YY Gem \citep{alef97}.
Positive detections  include the Algol system
V505 Sgr \citep{gunn99} and the pre-cataclysmic system 
V471 Tau \citep{patterson93, lim96d}. The former surprisingly
shows both a primary and a secondary eclipse although one of the components is
supposed to be inactive. The authors suggested that a radio coronal
component is located between the two stars, a conjecture that gives rise to
interesting theoretical models such as diffusion of magnetic fields from the 
active to the inactive star (on unreasonably long time scales, however), or radio emission from 
the  optically thin mass  accretion stream (which is found to be too weak), or field shearing by
the inactive companion. The radio emission of V471 Tau may originate from  
magnetic loops that extend to the white dwarf where they interact with its magnetic
field \citep{lim96d}.

Radio  rotational modulation is often  masked by intrinsic variability \citep{rucinski92}.
Well-documented examples are the RS CVn binaries CF Tuc \citep{budding99} for which 
\citet{gunn97} suggested the presence of material in the intrabinary region, and the RS CVn 
binary UX Ari, which appears to have radio-emitting  material concentrated 
above magnetic spots on the hemisphere of the K subgiant that is invisible from the companion 
\citep{neidhoefer93, elias95, trigilio98}. The clearest main-sequence example  
is AB Dor, with two emission peaks per rotation seen repeatedly over long time intervals.
Both peaks probably relate to preferred longitudes of active regions \citep{slee86, lim92, vilhu93}. 
The shape of the maxima suggests that the  radiation is intrinsically
directed in the source, which is plausible for synchrotron emission by 
ultrarelativistic electrons \citep{lim94}.

\subsection{Very Long Baseline Interferometry}\label{vlbi}

Interferometry at large baselines has long been a privilege of radio astronomy and
has proven versatile for numerous applications \citep{mutel96}. Various intercontinental
Very Long Baseline Interferometry (VLBI) networks have been arranged, and
dedicated networks such as the US Very Large Baseline Array (VLBA) or the UK MERLIN
network are routinely available. Apart from astrometric applications with sub-mas
accuracy \citep{lestrade90, lestrade93, lestrade95, lestrade99, guirado97}, VLBI 
provides coronal mapping at $\la 1$~mas angular resolution.

One of the principal, early VLBI results for RS CVn and Algol-like binaries relates to  
evidence for a compact core plus extended halo radio structure of a total size similar to the 
binary system size  \citep{mutel84, mutel85b, massi88, trigilio01}. Compact cores appear 
to be flaring sites, whereas halo emission corresponds to quiescent, low-level radiation, 
perhaps from decaying electrons from previous flares. There is evidence for non-concentric, 
moving, or expanding sources \citep{lestrade88, trigilio93, lebach99, franciosini99,
lestrade99}.

Pre-main sequence stars are attractive VLBI targets.
Although some sources are overresolved at VLBI baselines \citep{phillips93},
others have been recognized as components in close binaries such as HD 283447 
\citep{phillips96}; but most of the nearby weak-lined T Tau stars (see ``Evolution to 
the Main Sequence,'' below) reveal
radio structures as large as 10$R_*$, indicative of extended, global, 
probably  dipolar-like magnetospheres somewhat similar to those seen in RS CVn binaries 
\citep{phillips91, andre92}.  Perhaps also related to youth, magnetospheres
around Bp/Ap stars have also been observed with VLBI and have strongly supported global 
magnetospheric models (\citealt{phillips88, andre91}; see ``Stars at the Interface
Between Hot Winds and Coronae,'' below).

VLBI techniques have been  more demanding for single late-type dwarf stars, owing both
to lower flux levels and apparently smaller coronal sizes. Some  observations
with mas angular resolutions show unresolved quiescent or flaring sources, thus 
constraining the brightness temperature to  $T_b > 10^{10}$~K \citep{benz91, benz95}, whereas others
show evidence for extended coronae with coronal sizes up to several times the 
stellar size \citep{alef97, pestalozzi00}.

\begin{figure}[t!]
\hbox{
\hskip 0truecm\psfig{file=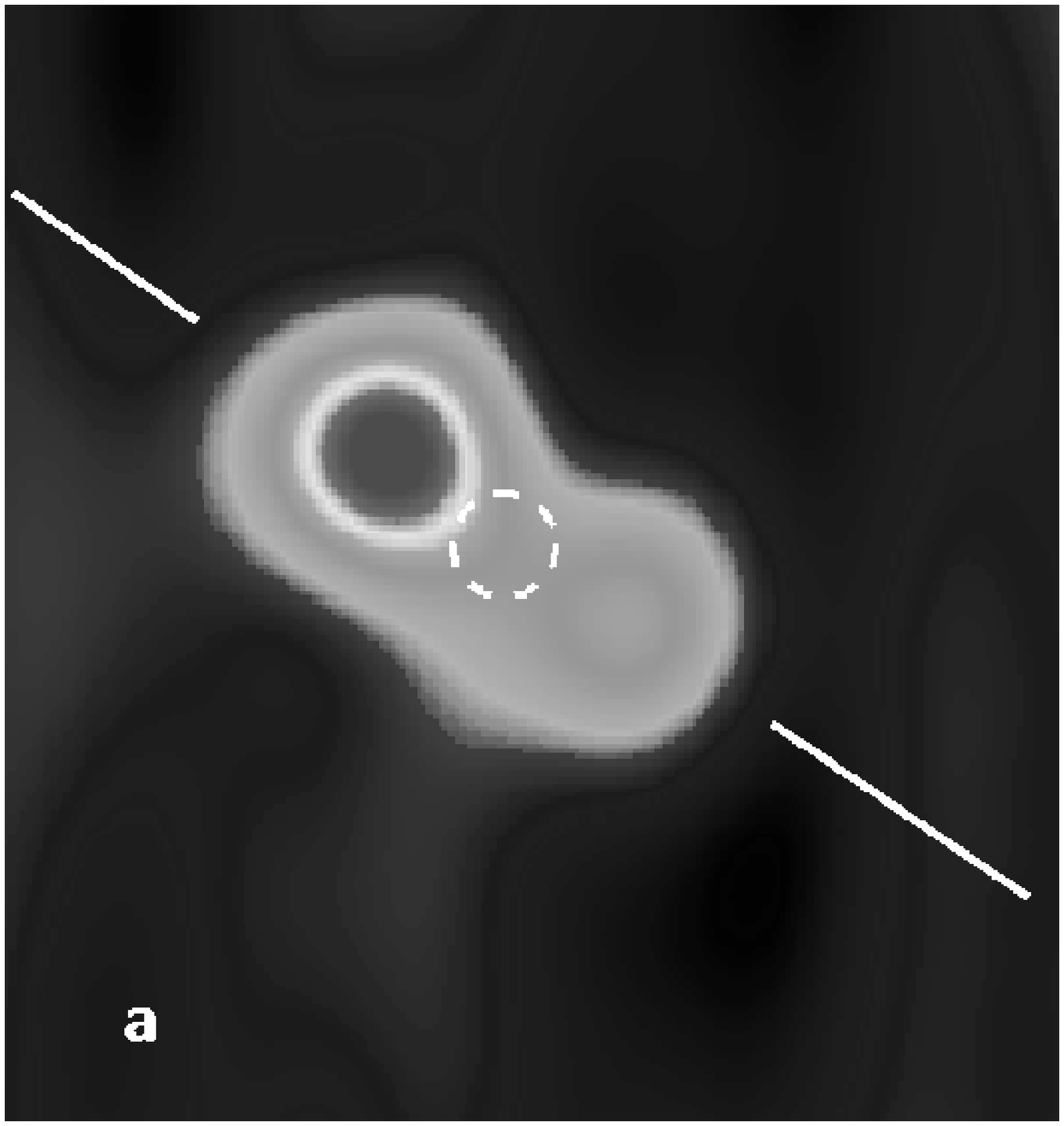,width=7.5truecm}
\hskip 0truecm\psfig{file=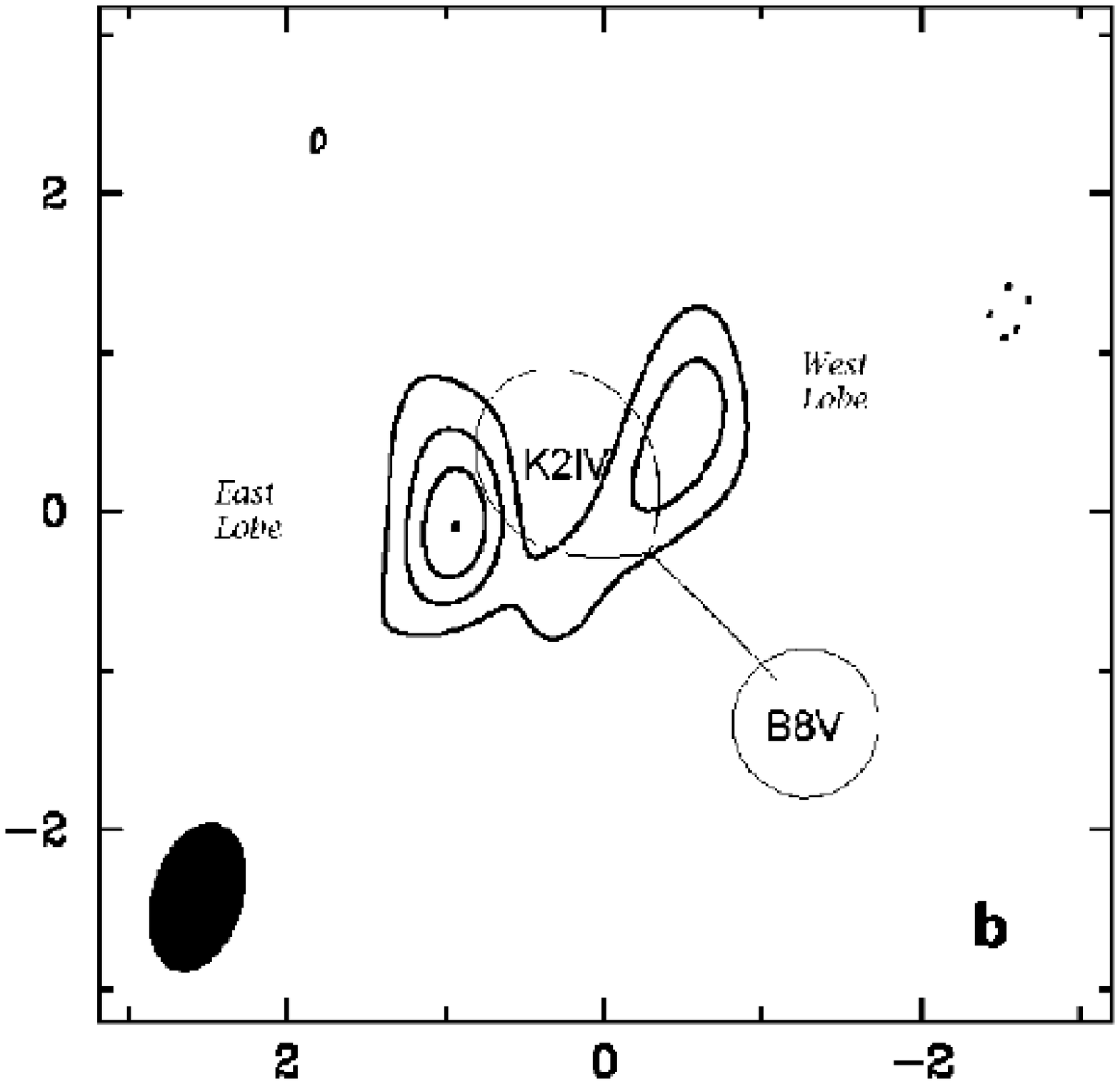,width=8.19truecm}
}
\caption{{\it (a)} VLBA image of the dMe star UV Cet; the two radio lobes
are separated by about 1.4~mas, while the best angular resolution reaches 
0.7~mas. The straight line shows the orientation of  the putative rotation 
axis, assumed to be parallel to the axis of the orbit of UV Cet around the 
nearby Gl 65~A. The small circle gives the photospheric diameter to size,
although the precise position is unknown (after Benz et al 1998). {\it (b)}
VLBA image of the Algol binary. The most likely configuration of the binary 
components is also drawn. The radio lobes show opposite
polarity (Mutel et al 1998; reproduced with permission of the AAS.)\label{vlba} }
\end{figure}

The dMe star UV Cet is surrounded by a pair of giant synchrotron lobes, with sizes up to 
$2.4\times 10^{10}$~cm and a separation of 4$-$5 stellar radii along the 
putative rotation axis of the star, suggesting very extended magnetic structures above the 
magnetic poles (\citealt{benz98}; Figure~\ref{vlba}a). VLBA imaging and polarimetry of Algol reveals 
a similar picture with two oppositely polarized radio lobes separated along a line
perpendicular to the 
orbital plane by more than the diameter of the K star (\citealt{mutel98}; Figures~\ref{vlba}b, 
\ref{models}b). 
Global polarization structure is also suggested in UX Ari, supporting the view that the 
magnetic fields are large-scale and well ordered \citep{beasley00}.

\subsection{Magnetospheric Models}\label{magnetosphere}

\begin{figure}[t!]
\hbox{\psfig{file=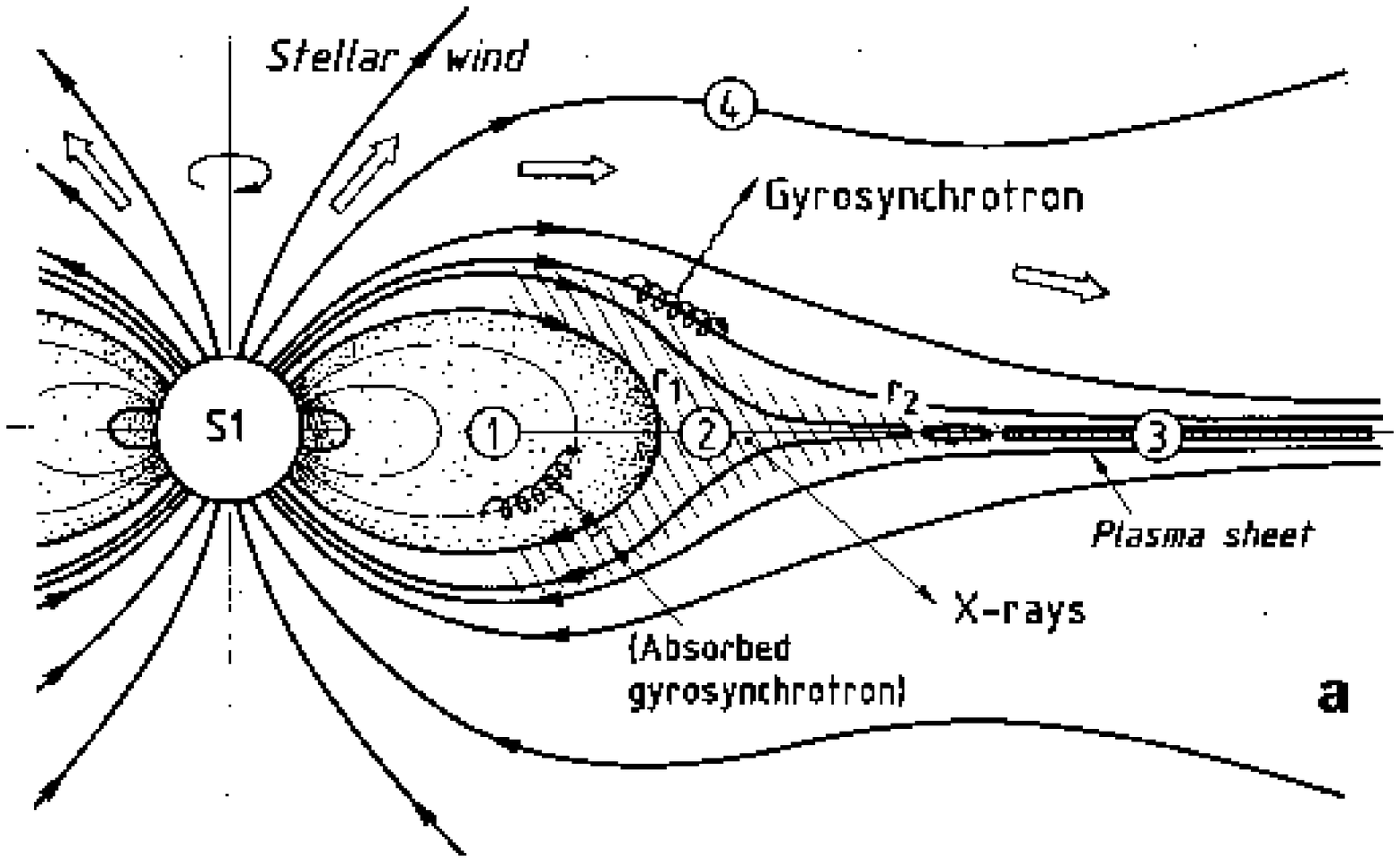,width=9truecm}
      \hskip 1truecm\psfig{file=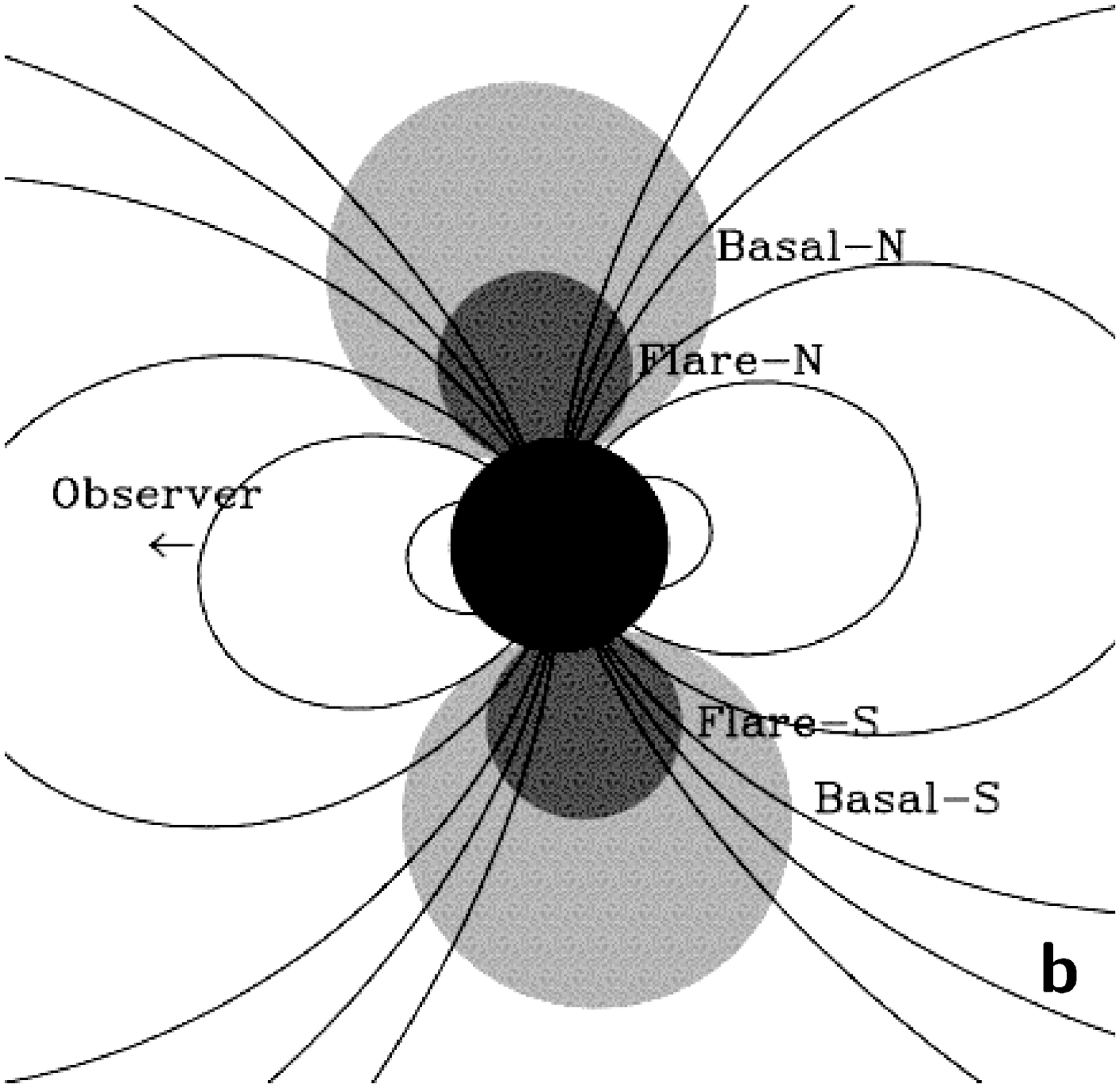,width=5.5truecm}
}
\caption{{\it (a)} Equatorial model for the magnetosphere of the young B star S1 in $\rho$ Oph
         \citep{andre88}. {\it (b)} Sketch for radio emission from a global dipole
	  consistent with the VLBI observation shown in Figure~\ref{vlba}b 
	  (\citealt{mutel98}; reproduced with permission of the AAS.)\label{models}}
\end{figure}

Quite detailed geometric models of large magnetospheres around RS CVn and Algol 
binaries, T Tau stars, and magnetic Bp/Ap stars have been designed based on
VLBI results, radio spectra and polarization. Common to all is a global, 
dipole-like structure somewhat
resembling the Earth's Van Allen belts (Figure~\ref{models}a).  
Stellar winds escaping  along magnetic fields draw the field
lines into a current sheet configuration in the equatorial plane.  
Particles accelerated in that 
region travel back and are trapped in the dipolar-like, equatorial magnetospheric 
cavity. Variants of this radiation belt model, partly based on theoretical work of 
\citet{havnes84}, have been applied to RS  CVn binaries \citep{slee87b, morris90, 
jones94, storey96}, in an optically thick version to Bp/Ap stars \citep{drake87b, 
linsky92} and in an optically thin version to a young B star  \citep{andre88}.

Such models are particularly well supported by polarization measurements in RS CVn 
binaries. For a given system, the polarization degree $p$ at any 
frequency is anticorrelated with the luminosity, while the sense of polarization
changes between lower and higher frequencies. For a stellar sample,  $p$ is inversely 
correlated with the stellar inclination angle such that low-inclination (``pole-on'') 
systems show the strongest polarization degrees \citep{mutel87, mutel98, morris90}.

When flares are in progress, the two-component core plus halo model appears to correctly 
describe radio spectral properties. Stronger magnetic fields  of 80$-$200~G are inferred
for the core and weaker fields of 10$-$30~G for the halo 
\citep{umana93, umana99, mutel98, trigilio01}.  
The frequency-dependent optical depth  makes the source small at high frequencies
(size $\approx R_*$, above 10~GHz) and large at small frequencies (size comparable
to the binary system size at 1.4~GHz; \citealt{klein87b, jones95}). This effect correctly 
explains the relatively flat, optically thick radio spectra seen during flares.

\section{STELLAR ROTATION, BINARITY, AND MAGNETIC CYCLES}

\subsection{Radio Emission and Stellar Rotation}

There is little doubt that rotation is largely responsible for the
level of magnetic activity in cool stars. It is clearly the young, rapidly rotating
stars or spun-up evolved stars in tidally interacting binaries that define the
vast majority of radio-emitting cool stars detected to date. The rather peculiar
class of FK Com stars, single giants with unusually short rotation periods of 
only a few days, are among the most vigorous emitters of gyrosynchrotron
emission \citep{hughes87, drake90, rucinski91, skinner91}, including extremely luminous
and long flares \citep{slee87b, bunton89}. However, large surveys of RS CVn binary systems 
find at best a weak correlation between radio luminosity and rotation parameters 
\citep{mutel85a, morris88, caillault88, drake89, drake92}, and it may 
even depend  on the optical luminosity class considered.  Although \citet{stewart88} 
reported a trend of the form $L_{\mathrm{R}} \propto R_*^{2.5}v^{2.5}$
($v$ is the equatorial velocity) for different luminosity classes, this
result was later challenged \citep{drake89} and may be related to using  peak
fluxes and flares. The absence of a correlation may be related
to the coronal saturation regime known in X-rays \citep{vilhu83, mutel85a, white89a}:
almost all radio-detected stars emit at the maximum possible and rotation-independent
X-ray level - hence perhaps also at the maximum possible radio level. There is obviously
a large potential for stellar radio astronomy in the less exotic regime below saturation! 
  
Moving to tighter binary systems in which the components are in (near-)contact, we would
expect  magnetic activity to increase or at least to stay constant. This is however  
not the case. The radio emission of such systems is significantly weaker than that of RS CVn 
binaries or active single stars, a trend that is also seen at other wavelengths 
\citep{hughes84, rucinski88, vilhu88}. 
Possible physical causes include i) a reduced differential rotation, hence a weaker dynamo
action \citep{beasley93b}, ii) a shallower convection zone in contact binary systems, and iii)
influence by the energy transfer between contacting stars \citep{rucinski95}.

\subsection{Activity Cycles}\label{cycle}

Radio polarization may be a telltale indicator for magnetic activity cycles on stars,
analogous to the solar 11~year cycle (or 22~years, if the magnetic polarity reversal is 
considered).  Long-term measurements in the Ca H\&K lines (the HK project; 
\citealt{baliunas98}) indicate the presence of activity cycles with durations of several 
years in low-activity stars,  but irregular long-term variations on active stars. In any case,
one would expect  reversals of the dominant sense of polarization to accompany any of these 
quasi-cycles. The contrary is 
true, to an embarrassing level: After decades of monitoring, many active stars show a 
constant sign of radio polarization throughout, both in quiescence and during flares \citep{gibson83,
white86, mutel87, kundushev88, jackson89, white95, mutel98} with few exceptions \citep{bastian87,
willson88, lang90}. This suggests the presence of some stable magnetospheric structures or 
a predominance  of strong magnetic fields in one polarity. A concerted
and ongoing effort to look for polarity reversals is negative at the time of this writing
(S. White, private communication).  

Perhaps the most convincing case yet reported in favor of a magnetic cycle is HR~1099 in
which the average radio  flux density correlates  with the spot coverage, revealing
a possible periodicity of 15$-$20 years \citep{umana95}. \citet{massi98} reported a surprisingly 
rapid quasi-periodicity in microwave
activity ($P \approx 56\pm 4$~days) and the accompanying sense of polarization
($P \approx 25$~days) in the RS CVn binary UX Ari. Whether such rapid oscillations
are related to an internal magnetic dynamo cycle remains to be shown.

\section{RADIO EMISSION FROM CHROMOSPHERES AND WINDS}\label{chromospheres}

\subsection{Lower Atmospheres of Cool Stars} 
         
\begin{figure}[t!]
\centerline{\psfig{file=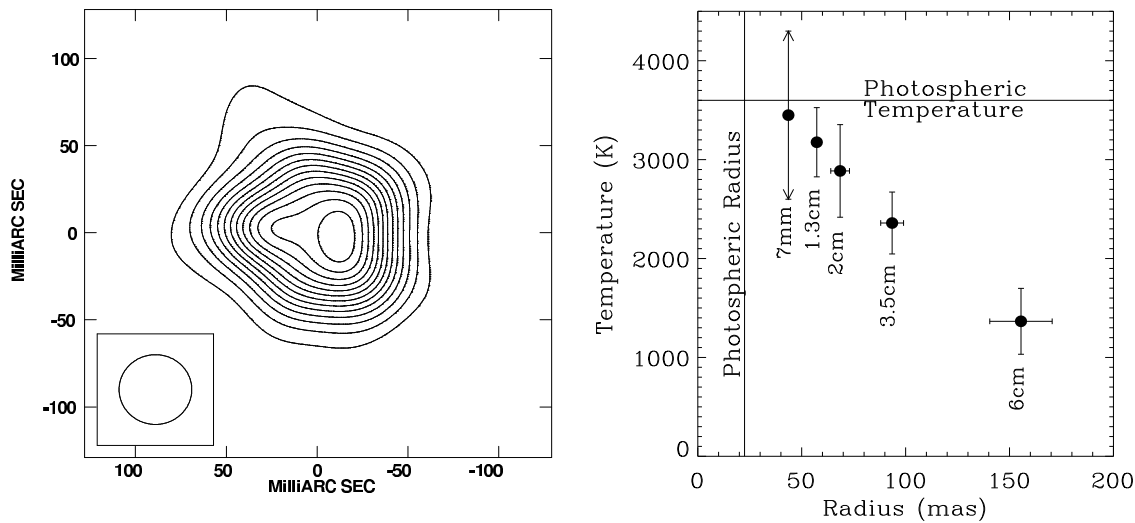,width=15truecm}}
\caption{{\it (a)} Betelgeuse observed at 43 GHz. The radio photosphere of this
star is resolved, with an angular resolution of 40~mas. {\it (b)} The atmospheric
temperature of Betelgeuse as a function of radius, observed at different
frequencies. (From \citealt{lim98}; reproduced with the permission of Nature.)\label{betelgeuse}}
\end{figure}

Stellar transition regions and chromospheres are expected to be 
radio sources as well (see ``Quiescent Emission from Coronae of Cool Stars,'' above). 
Among cool main-sequence stars,
the slightly evolved mid-F star Procyon is the only such source in the solar 
vicinity detected to date \citep{drake93}. In the red giant area, however, outer atmospheres
fundamentally change their characteristics.
Chromospheres  become as large as several stellar radii. Such sources are now well 
detected at radio wavelengths \citep{knapp95} and spatially resolved \citep{skinner97}, 
albeit with some surprises. Because their outer
atmospheres are optically thick at radio wavelengths, spatially resolved observations provide
a direct temperature measurement. \citet{reid97} inferred optically thick ``radio photospheres'' at 
about 2$R_*$ but at subphotospheric/nonchromospheric  temperatures.
Indeed, contrary to chromospheric UV measurements, the radio-derived temperature in the nearby supergiant
Betelgeuse is seen to drop systematically from optical-photospheric levels outward 
(\citealt{lim98}; Figure~\ref{betelgeuse}). 
This cool material  completely dominates the outer atmosphere. The authors suggested 
that cool, photospheric material is elevated in giant convection cells; dust formation in
this environment could then drive Betelgeuse's outflow.

\subsection{Winds from Cool Stars}

Could dwarf stars lose mass by a substantial stellar wind
\citep{doyle91, mullan92}? Such a wind, if (partially) ionized, would be a radio source. 
The very fact that coronal radio flares are seen at GHz frequencies implies that the extended
winds must be optically thin down to coronal heights, and this places stringent limits on 
the mass loss rate. Sensitive millimeter 
measurements and theoretical arguments have constrained such mass loss for M dwarfs
to $\la 10^{-13} M_{\odot}$~yr$^{-1}$ or $\la 10^{-12} M_{\odot}$~yr$^{-1}$  for a $10^4$~K 
or a $10^6$~K wind, respectively \citep{lim96c, lim96a, vdoord97}, in agreement with
upper limits derived from UV observations \citep{wood01}. 
Stringent upper limits to the radio emission of solar analogs also help confine the 
mass loss history of the Sun or solar-like stars \citep{drake93, gaidos00}.
An upper limit to the mass loss during the Sun's main-sequence life of  $6\%$ indicates that the 
Sun was never sufficiently luminous to explain the ``Young Sun Paradox'', i.e., the 
suggestion that the young Sun was  more luminous in early times given the apparently
much warmer climate on Mars \citep{gaidos00}.

Winds become progressively more important toward the red giants, in particular beyond the 
corona-cool wind dividing line.   Ionized-mass losses between 
$10^{-10}-10^{-9}M_{\odot}$~yr$^{-1}$ are indicated, increasing toward cooler and more
luminous stars \citep{drake86b}. However, the surface mass loss flux is  
similar in all stars, in fact also similar to the solar wind mass loss flux. The 
coolest, late-M and C (super)giants still support massive winds, but the much weaker 
radio emission  indicates that the ionization fraction drops by at least an order of 
magnitude compared to earlier M giants, i.e., their chromospheres must be cool, and 
their optical depth at radio wavelengths becomes small \citep{drake87a, drake91a, luttermoser92}. 
The same holds for technetium-deficient S and M-S giants, although white dwarf 
companions of some of them may sufficiently ionize the outer atmosphere of the giant to become 
visible at radio wavelengths \citep{drake91b}. This latter mechanism is probably also 
relevant in the interacting ``symbiotic binaries'' that usually consist of a red giant and a 
white dwarf companion and that are radio sources with considerable luminosities. The white dwarf 
is sufficiently UV strong to ionize part of the cool-star wind \citep{seaquist84,seaquist93, taylor84,
seaquist90}. No appreciable radio emission is, however, detected from the similar class of G 
and K-type Barium stars (showing overabundances of Ba and other nuclear-processed elements),
although they show evidence of  white-dwarf companions \citep{drake87c}.

\subsection{Ionized Winds and Synchrotron Emission from Hot Stars}

Because OB and Wolf-Rayet stars shed strong ionized winds, they emit  thermal-wind radio 
emission (equations~\ref{wind}). Some of the hot-star radio 
sources are very large and resolved by the VLA because the optically thick surface
(equation~\ref{windsurface}) is located at hundreds of stellar radii. Under the assumption
of a steady, spherically symmetric wind, wind mass-loss rates
of $\dot{M} \approx 10^{-6}-10^{-5}~M_{\odot}$~yr$^{-1}$ are inferred for O and B stars
\citep{scuderi98}, and $\dot{M} \approx 2-4\times 10^{-5}~M_{\odot}$~yr$^{-1}$ for 
WR stars \citep{bieging82, leitherer95, leitherer97}. The inferred mass-loss rate is closely correlated 
with the bolometric luminosity, $\mathrm{log}\dot{M} = (1.15\pm 0.2)\mathrm{log}L + C$
\citep{scuderi98}, in good agreement with H$\alpha$ measurements. In the case 
of colliding wind binaries, the thermal radio emission can be further enhanced by contributions 
from the wind-shock zone \citep{stevens95}. The radio emission level drops appreciably toward 
intermediate spectral classes of B, A, and F, probably owing to a steep
decrease of the ionized mass loss \citep{drake89a}.  

It came as  a surprise when several OB stars \citep{abbott84, bieging89a, drake90a, contreras96}
and WR stars  \citep{becker85, caillault85, abbott86, churchwell92, chapman99}  were 
found to show non-thermal, synchrotron-like radio spectra, some of
them associated with  short-term variability \citep{persi90}.   
Up to 50\% of the  WR stars appear to be non-thermal sources, and this
fraction is half as large for OB stars \citep{leitherer97, chapman99, dougherty00a}. 
Hot stars are not thought to produce magnetic fields via a dynamo. Moreover, because the
wind is optically thick to radio emission out to hundreds of stellar radii, the non-thermal
component must originate at such large distances as well. A coronal model
should therefore not apply although highly variable radio emission and a poor correlation with X-ray
behavior may suggest that magnetic fields play a role in the structuring of the winds 
\citep{waldron98}. Viable alternatives include 
synchrotron emission from electrons accelerated in shocks of unstable winds of single
stars \citep{white85, caillault85}, in colliding-wind shocks in massive binaries \citep{eichler93},
and in the interaction zone between the thermal wind and a 
previously ejected shell (Luminous Blue Variable phase; \citealt{leitherer97}). Magnetic fields inferred 
for the photospheric level are of order 1 Gauss \citep{bieging89a, phillips90}, and up to  50~G in
synchrotron envelope models \citep{miralles94}.

\begin{figure}[t!]
\centerline{\psfig{file=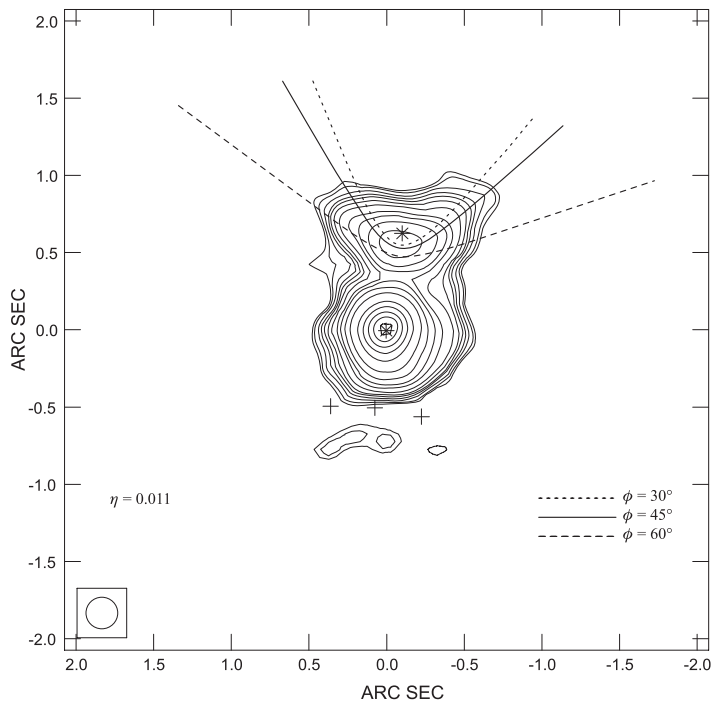,width=10truecm}}
\caption{VLA map of the WR 147 system observed at 3.6~cm. 
   Also shown are calculated model curves that follow the shock formed 
   by the two-wind interaction (from \citealt{contreras99}; reproduced with
   the permission of the AAS).\label{wr147}}
\end{figure}

Shocks are attractive features as they can accelerate electrons by the first-order Fermi 
mechanism, with a predicted electron energy power-law index $\delta = 2$ \citep{bell78}.
\citet{skinner99},
however, found a significant deviation from this model for a colliding-wind binary
but successfully interpreted their radio spectra with an absorbed synchrotron spectrum 
from a quasi-monoenergetic electron population. The origin and stability of such a population are 
unclear. The colliding-wind model  found strong support when it became evident that
most non-thermal sources are indeed binaries \citep{dougherty00a} and that
the non-thermal sources are located between the two stars, separate
from the wind  \citep{moran89, churchwell92, dougherty96, dougherty00b, williams97, 
chapman99}.  Convincing evidence for colliding winds is available for WR~146 
\citep{dougherty00b}, Cygnus OB2 No 5 \citep{contreras97},
 and WR~147 that all  show a thermal source 
coincident with the WR star and a separate bow-shock shaped non-thermal source in
the wind collision zone close  to a lower-mass companion (Figure~\ref{wr147}, 
\citealt{contreras99}).  If the stars are in  a strongly eccentric orbit,
the wind-shock zone between two stellar components can enter the thermal-wind 
radio photosphere and become absorbed, thus producing strong
long-term variability of the radio flux  \citep{williams90, williams92, williams94, 
white_rl95, setia00}.  Further long-term synchrotron variability may be caused
by long-term modulation in the magnetic field  along the orbit, whereas short-term
($\sim$daily) variability may be due to clumps in the wind that arrive at the shock
\citep{setia00}.

The rather inhomogeneous class  of Be stars shows evidence for very large ($\sim 100R_*$)
high-density outflowing disks that have been probed at radio wavelengths. Steeply increasing
radio spectra and some flux variability are characteristic, but there is considerable
debate on the source geometry \citep{taylor87, taylor90, drake90a, dougherty91}.

There have been a number of interferometric observations of non-thermal sources among hot 
stars. Source sizes of order $100R_*$, exceeding the size of the  optically thick
surface of the thermal wind, have been measured. They imply brightness temperatures 
up to at least $4\times 10^7$~K and thus further support the non-thermal interpretation
\citep{felli89, felli91a, felli91, phillips90}.

\section{STAR FORMATION AND THE SOLAR-STELLAR CONNECTION}

\subsection{Radio Emission from Low-Mass Young Stellar Objects}

Large numbers of visible T Tau stars and  embedded infrared sources
in star forming regions are strong radio sources \citep{garay87, felli93}. 
Schematically, pre-main sequence evolution is thought to proceed through four consecutive
stages with progressive clearing of circumstellar material \citep{shu87}:
i) Cold condensations of infalling molecular matter, forming a hydrostatic 
low-luminosity protostellar object, with the bulk mass still accreting (``Class 0 source''). 
ii) Formation of a deeply embedded  protostar 
(``Class I source'') through  further accretion via a massive accretion disk, associated 
with strong polar outflows. iii) The ``classical T Tau'' (CTT, ``Class II'') phase with an
optically visible star accompanied by a thick accretion disk, a weak outflow, and 
possibly a weakly ionized wind; and iv) the ``weak-lined T Tau'' (WTT, ``Class III'') phase 
at which disk and circumstellar material have largely been dissipated, and the star approaches the
main sequence. The evolutionary sequence  is somewhat controversial (see below)  
and may in fact describe the evolutionary phase of the circumstellar material rather 
than of the star itself. 

At first inspection, and especially in the outer reaches of
the molecular clouds such as the $\rho$ Oph cloud \citep{andre87, andre88, stine88, 
magazzu99}, one encounters predominantly  WTTS that appear to have evolved past the CTT 
phase. How early in its infancy can a star be to develop strong radio signatures? Whereas deeper 
surveys of $\rho$ Oph have accessed several deeply embedded infrared sources with high radio 
luminosities  (log$L_{\mathrm{R}} > 15$, some of which are class I sources, \citealt{andre87,
brown87, leous91, feigelson98}) many of the  radio-strong WTTS in the $\rho$ Oph dark cloud 
are significantly closer to the sites of current star formation, and therefore younger,
than the typical radio-weaker WTTS and CTTS. It seems that radio-strong WTTS evolve
directly from embedded protostars. \citet{andre92} speculated that in some cases strong fossil 
magnetic fields accelerate both dissipation of circumstellar material  and
formation of large magnetospheric structures on short time scales of 
$\sim 10^6$~years.

\begin{figure}[t!]
\centerline{\psfig{file=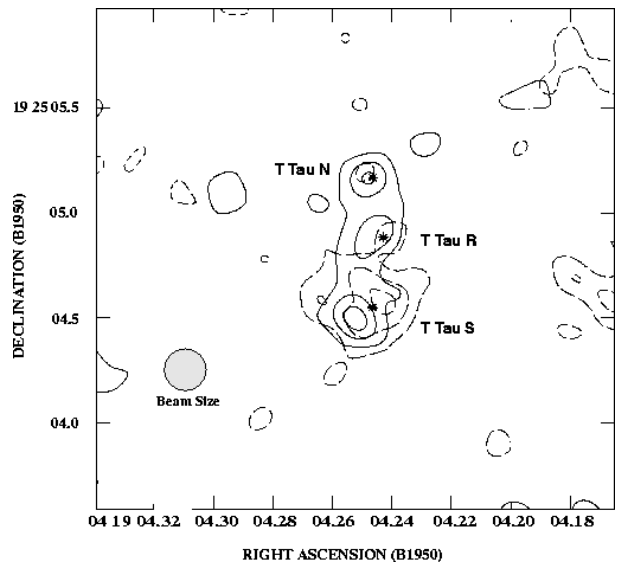,width=8truecm}}
\caption{Observation of the T Tau system at 6~cm with the MERLIN interferometer. 
Right- and left-hand circularly polarized components are shown dashed and solid,
respectively. The offset of the polarized flux centroids in T Tau(S) is
interpreted in terms of polarized outflows (Ray et al. 1997; reproduced with 
the permission of Nature.)\label{ttau}}
\end{figure}

Genuine, embedded class I protostars have most often been detected as thermal sources,
and this emission is predominantly due to collimated thermal winds or
jets. These jets are probably ionized by neutral winds that collide with the ambient medium at distances
of around 10~AU and that are aligned with molecular outflows \citep{bieging85, snell85, 
brown87, curiel87, curiel89, curiel90, rodriguez89, rodriguez90, rodriguez95, morgan90,
marti93, garay96, suters96, torrelles97}.
In the case of more massive stars, the radio emission can also originate from optically thick
or thin  compact HII regions  (e.g., \citealt{hughes88, estalella91, gomez00}), 
or from ionized winds \citep{felli98};
even (thermal or non-thermal) gyrosynchrotron emission has been proposed
given the high brightness temperatures, small sizes, variability,  and negative spectral 
indices of some sources \citep{hughes91, hughes95, garay96}.
For a review of thermal radio jets driving outflows and Herbig-Haro objects, see,
e.g., \citet{anglada95, anglada96}, and \citet{anglada98}. Ionized circumstellar material 
and winds easily become 
optically thick and therefore occult any non-thermal, magnetic emission from close to the 
star. However, the discovery of polarization in T Tau(S) \citep{phillips93}, in IRS~5 
\citep{feigelson98}, in protostellar jet sources \citep{yusef90} and the jet outflows
themselves (\citealt{curiel93, hughes97, ray97}; Figure~\ref{ttau}),
as well as variability and negative spectral indices in T Tau(S) \citep{skinner94}  
provided definitive evidence for magnetic fields and particle acceleration in these  
class I objects.

Moving toward the youngest accreting or class 0 sources, centimetric thermal radio 
detections probably again relate to jets/collimated winds that drive massive outflows, whereas 
the bulk of the
emitted power leaves the system at mm/submm wavelengths from cold ($\la$20~K) dust,
a defining property of class 0 protostars \citep{andre93}. Several objects have been
detected, with ages of the order of only $10^4$~years and central stellar masses of only 
$\approx 0.05M_{\odot}$ \citep{leous91, andre93, bontemps95, yun96, andre99, gibb99, reipurth99}, 
marking the very beginning of the protostellar accretion phase. Radio emission is thus 
a sensitive tracer for the presence of an embedded, nascent but already formed protostellar 
core as opposed to a contracting cloud fragment  (see, e.g., \citealt{yun96}).

\subsection{Evolution to the Main Sequence}\label{pms}

Early VLA surveys of CTTS quickly recognized their thermal wind-type
emission with rising spectra and large angular sizes \citep{cohen82, bieging84, cohen86,
schwartz84, schwartz86}.  
The enormous kinetic wind energy derived under the assumption of a uniform spherical wind 
suggests anisotropic 
outflows while structural changes in the radio sources indicate variable outflows, probably
along jet-like features \citep{cohen86}. Mass loss rates are estimated to be 
$\la 10^{-7}M_{\odot}$~yr$^{-1}$ \citep{andre87}. It is important to note that the thermal 
radio emission says nothing about  the presence or absence of stellar magnetic fields. 
CTTS do show a number of magnetically induced phenomena, but
whatever the possible accompanying radio emission, it is thought to be absorbed by the
circumstellar ionized wind, unless huge magnetospheric structures reach beyond the optically
thick wind surface \citep{stine88, andre92}.

The CTT stage is pivotal for planetary system formation as massive accretion
disks are present. Their dust emission dominates the systemic luminosity at 
millimeter wavelengths. Large ($>$1000~AU) molecular and dust features have been 
mapped at such wavelengths; however, observations of the solar-system sized inner disks (100~AU)
have been challenging and require high angular resolution. At cm wavelengths,
thermal collimated outflows may become  dominant over dust disks but both features 
can be mapped simultaneously in some cases, revealing two orthogonal structures, one 
a jet and one an edge-on disk \citep{rodriguez92, rodriguez94, wilner96}.
 
At ages of typically $1-20\times 10^6$~years, most T Tau stars  dissipate their
accretion disks and circumstellar material and become similar to main-sequence   
stars albeit  at much higher magnetic activity levels, probably induced by 
their short rotation periods \citep{oneal90, white92a}. The presence of
huge flares \citep{feigelson85, stine88, stine98}, longer-term
variability,  and falling spectra clearly point to non-thermal gyrosynchrotron emission
\citep{bieging84, kutner86, bieging89b, white92a, felli93, phillips96} analogous to more evolved
active stars.  Conclusive radio evidence for the presence of solar-like magnetic fields 
in WTTS came with the detection of weak circular polarization during flare-like modulations, 
but also in quiescence \citep{white92b, andre92, skinner93a}. 
Zeeman measurements confirm the presence of kGauss magnetic fields
on the surface of some of these stars (e.g., \citealt{johns99}).
Extreme particle energies 
radiating synchrotron emission may be involved, giving rise to  linear polarization in 
flares on the WTT star HD 283447 \citep{phillips96}. VLBI observations showing large 
($\sim$10$R_*$) magnetospheric structures with brightness temperatures of
$T_b \approx 10^9$~K fully support the  non-thermal picture \citep{phillips91}.
 
As a WTT star  ages, its radio emission drops rapidly on time scales of a few Million years  
from luminosities as high as $10^{18}$erg~s$^{-1}$Hz$^{-1}$ to values around or below
$10^{15}$erg~s$^{-1}$Hz$^{-1}$ at ages beyond 10~Myr. Young age of a star
is thus favorable for strong radio emission  \citep{oneal90, white92a, chiang96}, whereas 
toward the subsequent Zero-Age Main-Sequence (ZAMS) stage it is only the very rapid 
rotators that keep producing radio emission at a $10^{15}$erg~s$^{-1}$Hz$^{-1}$ 
level \citep{carkner97, magazzu99, mamajek99}.

\subsection{Low-Mass Stars on the Main Sequence}

The pace at which radio emission decays toward the ZAMS  has 
prevented systematic radio detections of cool main-sequence  stars beyond 10$-$20~pc. Much
of the evolutionary trends known to date have been derived from fairly
small and very select samples of extraordinary stars. Although many nearby
M dwarfs are probably quite young and located near the 
ZAMS, their ages are often not well known. An interesting exception is
the proper-motion dM4e companion to AB Dor, Rst137B, detected as a surprisingly
luminous steady and flaring radio star \citep{lim93, beasley93a}. With an age of $\sim 5-8 \times
10^7$~years, it may indicate that stars approaching the main sequence go through 
a regime of strongly enhanced magnetic activity. The AB Dor pair is a member of the 
Local Association, also known as the Pleiades Moving Group, a star stream of an age 
($\sim$50$-$100~Million years) 
corresponding to near-ZAMS for F$-$K stars. A handful of further stream  members
with vigorous radio emission  are now known in the solar  neighborhood, most notably 
PZ Tel (cited in \citealt{lim95}), HD 197890 (K1~V; \citealt{robinson94}), 
HD~82558 (K1~V, \citealt{drake90}), EK Dra (G0~V, \citealt{guedel94b, guedel95a}), and 47 Cas 
(F0~V + G~V; \citealt{guedel95c, guedel98}).

These objects are analogs of young open cluster stars with the observational advantage 
of  being much closer. Clusters, however, are preferred when more precise stellar ages or large
statistics are required. Some young  clusters  house a select group of 
ultra-fast rotators, stars at the extreme of dynamo operation with rotation periods   
$\la 1$~day. The absence of any detections in Bastian et al.'s (1988) Pleiades radio survey
suggests that their flare properties are similar to solar neighborhood stars rather than to
the much more energetic outbursts occasionally seen in star forming regions. 
Flaring or quiescent radio emissions have been detected from G$-$K-type 
members  of both the Pleiades \citep{lim95} and the Hyades \citep{white93, guedel96b}
although these examples, in part ultra-fast rotators at saturated activity levels
and in part binary systems, clearly represent only the extreme upper end of magnetic activity. 
Radio emission of normal, single solar analogs rapidly declines to undetectable 
levels after a few hundred Million years \citep{gaidos00}.

\subsection{Herbig Ae/Be stars}

The evolution of intermediate-mass ($3-20M_{\odot}$) stars is quite 
different from, and much faster than, that of low-mass stars as they still accrete  while 
already on the main sequence.
Given their intermediate spectral range, it is of great interest to know whether
Herbig Ae/Be pre-main sequence stars support convective, magnetic dynamos or whether 
they resemble more massive wind sources. Radio emission is the ideal discriminator. 
A wind-mass loss interpretation is compatible with the expected mass loss rates of 
$10^{-6}-10^{-8}M_{\odot}$yr$^{-1}$
\citep{guedelcatala89, skinner93b}. This interpretation is supported by
the large radio sizes (order of 1$^{\prime\prime}$) and the absence of circular polarization or
strong variability. The radio luminosity is also correlated with the stellar temperature
and bolometric luminosity. This is expected because  wind mass-loss rates increase toward
higher stellar masses \citep{skinner93b}. New radio observations complemented
by millimeter measurements further indicate the presence of substantial dust envelopes 
\citep{difrancesco97}. 

Non-thermal gyrosynchrotron sources exist as well among Herbig stars, although they
are the exception \citep{skinner93b}. The evidence comes primarily from negative
spectral indices, including the extremely X-ray strong proto-Herbig star EC95 = S68-2
that further supports a coronal model based on its $L_{\mathrm{X}}/L_{\mathrm{R}}$ 
ratio \citep{smith99}.

\section{STARS AT THE INTERFACE BETWEEN HOT WINDS AND CORONAE}\label{bpap} 

Owing to missing convection, no dynamo-generated magnetic fields are expected on stars
earlier than spectral type F, nor should there be massive  ionized winds on main-sequence
stars of spectral type B and later, given the weak radiation 
pressure. Indeed, few detections have been reported as early as
spectral type F0, although some have been found at quite high luminosities compatible with
the gyrosynchrotron mechanism, including
main-sequence candidates \citep{guedel95b} and the supergiant $\alpha$ Car
\citep{slee95}. \citet{brown90} surveyed a number of normal A-type stars and found stringent 
upper limits to any radio emission and thus to the mass-loss rate - very much
in agreement with expectations. However, as it turns out, this spectral range is shared by
some of the more provocative radio detections.

Many ``magnetic chemically peculiar''  Bp/Ap stars maintain considerable radio 
emission (log$L_{\mathrm{R}}$ $\approx 15-18$; \citealt{drake87b}), including very young 
objects such as S1 in $\rho$ Oph \citep{andre88}.   They all relate to the hotter 
(O9$-$A0 spectral type) He-strong and He-weak/Si-strong classes, 
whereas the cooler (A type) SrCrEu peculiarity classes remain undetected 
\citep{drake87b, willson88, linsky92, leone94}. Similarly,   
``non-magnetic''  Am and  HgMn stars remain undetected despite recent claims
that these stars may have magnetic fields as well \citep{drake94}. The emission mechanism
is likely to be gyrosynchrotron as judged from flat spectra \citep{linsky92, leone96},
high brightness temperature \citep{phillips88}, variability, and sometimes moderate
polarization \citep{linsky92}.

The presence of kilogauss magnetic fields on magnetic chemically peculiar stars 
has been known since 1947 \citep{babcock47}. The fields are generally assumed to be of
global, dipolar topology. There is little photospheric motion that could stir magnetic 
footpoints, but weak winds could  draw the 
magnetic field lines into an equatorial current sheet, thus producing a global
``van Allen Belt'' magnetospheric structure as described in ``Magnetospheric Models,''
above.
Estimated non-thermal source radii are a few stellar radii, confirmed by VLBI 
observations \citep{phillips88, andre91} that also conclusively established the 
non-thermal nature of the radio emission, with brightness temperatures of
$T_b \approx 10^8-10^9$~K.

The wind-controlled magnetospheric model is  further supported qualitatively by a parameter
 dependence of  the form $L_{\mathrm{R}} \propto 
 \dot{M}^{0.38}B_{\mathrm{rms}}^{1.06}P_{\mathrm{rot}}^{-0.32}$
where $\dot{M}$ is the  estimated wind mass-loss rate for the spectral type,  
$B_{\mathrm{rms}}$ is the  root-mean square value of the
longitudinal magnetic field, and $P_{\mathrm{rot}}$ is
the stellar rotation period (the dependence on the latter is marginal; \citealt{linsky92}).
Evidence for rotational modulation probably due to field misalignment has been 
found by \citet{leone91, leone93}, and \citet{lim96b}, the latter authors 
reporting modulation of the polarization degree and sign. A surprise detection was the 
phase-dependent 100\% polarized radio emission from a Bp star that suggests a strongly beamed,
continuously radiating electron cyclotron  maser \citep{trigilio00}. 

A binary class in this spectral range that is thought to be intermediate in evolution 
between the young semi-detached B-type $\beta$ Lyr system that shows radio evidence for large 
systemic wind mass loss \citep{umana00}, and evolved normal  Algol binaries with
non-thermal coronal emission is defined by
the rather inhomogeneous sample of Serpentid stars or ``Peculiar Emission Line Algols''
(PELAs). These typically consist of an A$-$B type primary and a F$-$K type companion
with strong mass transfer into a geometrically thick accretion disk around the 
early-type star, and a common, thin envelope.  They have shown an appreciable level of 
radio emission, first thought to be gyrosynchrotron emission based on their large luminosities
\citep{elias90} but later suspected to be wind sources given their spectral indices
close to the standard wind law \citep{elias93}.

\section{RADIO EMISSION FROM BROWN DWARFS}

Brown dwarf  stars have masses below the critical mass of $\approx 0.08M_{\odot}$ 
required for  stable hydrogen burning in the stellar core. 
Detecting radio emission that shows variability, polarization, or gyrosynchrotron-like
spectra from brown dwarfs would provide very strong evidence for the existence of surface magnetic fields.
However, after brown dwarfs cease to burn deuterium, the internal convection decreases,
probably diminishing the generation of magnetic fields \citep{neuhauser99}. A deep 3.6~cm 
survey with the  VLA, concentrating on older targets,  did indeed fail to produce radio  detections 
\citep{krishnamurthi99}. 
One young candidate brown dwarf has been reported with strongly variable radio emission
(P. Andr\'e, cited in \citealt{wilking99} and \citealt{neuhauser99}). But again,
nature offered more riddles: The first radio-detected bona-fide brown dwarf, LP944-20, is in fact quite
old ($\sim$500~Million years), but showed both  flaring and quiescent episodes 
\citep{berger01}.  Its radio 
emission is orders of magnitude higher than  would be expected from the active-stellar 
relation~(\ref{lxlrrelation}). In any case, the observed flaring radio emission is a 
telltale signature of magnetic fields, and proves that solar-like coronal activity is present
in substellar objects - even old ones.

A promising method for detecting sub-stellar objects with sufficiently strong magnetic fields
involves the operation of the cyclotron maser. In fact, Jupiter's decametric
radiation can be as strong as solar coherent bursts, although at somewhat longer
wavelengths. Searching nearby stars for low-frequency coherent bursts may, by positional
analysis, reveal  companion planets or brown dwarfs.
Such a study was performed by \citet{winglee86}, albeit with null results. We 
note in passing that one of their targets, Gl~411/Lalande 21185, has in the meantime been
shown by astrometric means to possess at least one planetary companion \citep{gatewood96}.

\section{STELLAR ASTROMETRY}\label{astrometry}
         
 VLBI techniques have been used for astrometric purposes, i.e., measurements of
positions, proper motions, binary orbital motions, and parallaxes, to 
an accuracy of fractions of a milliarcsecond \citep{lestrade90, lestrade93, lestrade95,
lestrade99}. While one fundamental purpose of such programs is to establish the astrometric link 
between the radio and the optical coordinate reference frames  that is eventually of great 
importance for numerous dynamical studies, invaluable astrophysical spin-offs were 
obtained. Astrometry of the Algol triple system (a 2.9-day period  A7m+K0~IV  
close binary with mass exchange, in a wide 680~d orbit around a B8~V star) showed unequivocally
that the radio emission is related to the magnetically active K subgiant.  The most
startling result (confirming earlier, indirect optical polarization measurements) was 
that the planes of the short and the long orbits are orthogonal to each other 
\citep{lestrade93, lestrade99}, a challenge for stability theories. Further
important highlights include 
i) precise  measurements of distances through parallaxes, including distances to 
    star forming regions, 
ii) identification of binarity or multiple systems, and
iii) rapidly moving ejecta after a stellar flare (see Lestrade  et al. 1999 for details).
A dedicated pre-Hipparcos VLBI study of the AB Doradus system 
solved for the accurate  distance (moving the star to 2/3 its previously assumed distance 
and placing it on the ZAMS)  and for a short-period disturbance attributed to
the presence of a low-mass object (possibly a brown dwarf) in orbit around this star \citep{guirado97}.    
         
\section{SUMMARY AND CONCLUSIONS}

Stellar radio astronomy has matured over the past few decades to a science that
is  indispensable for our understanding of stellar atmospheres. Historical 
milestones include, among many others, the discovery of steady and flaring non-thermal 
and polarized emission in cool stars, testifying to the importance of 
highly energetic processes; the recognition that these phenomena are ubiquitous
in many classes of convective-envelope stars; observations of very large, apparently stable
magnetospheric structures, unlike anything known from the Sun, around various types
of magnetically active stars such as T Tau stars, Bp stars, dMe stars, or RS CVn binaries;
the discovery of non-thermal emission produced in (wind-collision) shocks of hot-star
atmospheres; gyromagnetic and flaring emission from deeply embedded protostellar objects,
testifying to the importance of magnetic fields back to the earliest moments of a stellar
life; and flaring radio emission from sub-stellar objects not previously thought to support
stellar-like convective outer envelopes. Radio methodology has become a standard to estimate
magnetic fields in cool stars, to determine mass loss in stars with ionized winds, to
spatially resolve and map structures at the milliarcsecond level, and to simply prove the
presence of magnetic fields through polarization measurements. 

Far  from being an auxiliary science to research at other wavelengths, stellar radio
astronomy should prepare to address outstanding problems to which it has unique access,
although more sensitive instruments are needed. Questions of particular interest include:
Are there relevant high-energy processes and magnetic fields in class 0 protostars?
Are accretion processes important for the high-energy mechanisms and the generation
of large-scale magnetic fields? Are there magnetic fields in hot stars, and what role 
do they play in the  winds? Are brown dwarfs usually quiescent radio emitters? 
Do they maintain stable magnetic fields? What is the structure of their coronae?
Are there intra-binary magnetic fields in close binary stars? How do large magnetospheres
couple to the more compact X-ray coronae? Are quiescent coronae fed by numerous 
(micro-)flares?  
\vskip 0.5truecm

Acknowledgements: It is a pleasure to thank Marc Audard, Arnold Benz, Stephen Skinner, Kester Smith, and Stephen 
   White for helpful discussions, and  Philippe Andr\'e, Tim Bastian, Arnold Benz,  Maria Contreras,
   Jeremy Lim, Robert Mutel, Tom Ray, Luis Rodr\'{\i}guez, and Stephen White for providing figure 
    material. The introductory Einstein quotation is cited from "Einstein sagt", ed A Calaprice and
   A Ehlers (1997, Munich: Piper) and from "The Expanded Quotable Einstein", ed A Calaprice
   (2000, Princeton: Princeton University Press).


\newpage

\listoffigures

\end{document}